\documentclass[preprint,11pt]{elsarticle}
\usepackage[margin=2.5cm]{geometry}
\usepackage{graphicx}
\usepackage{mathtools}
\usepackage{booktabs,multirow,array}
\usepackage{makecell}
\usepackage{subfig}
\usepackage{rotating}
\usepackage{url}

\usepackage[dvipsnames]{xcolor}
\usepackage{tabularx}
\usepackage{adjustbox}
\usepackage{diagbox}
\usepackage{makecell}
\usepackage{lineno}
\usepackage{setspace}
\usepackage{enumitem}
\usepackage{cite}

\usepackage[title]{appendix}
\usepackage{chngcntr}

\usepackage{color,soul}
\usepackage{textcomp}

\usepackage[hidelinks, colorlinks=true,linkcolor=blue]{hyperref}

\begin{document}
	\begin{frontmatter}

		\title{Machine Learning Framework for High-Resolution Air Temperature Downscaling Using LiDAR-Derived Urban Morphological Features}

		\author{Fatemeh Chajaei, Hossein Bagheri}
		\address{Faculty of Civil Engineering and Transportation, University of Isfahan, Isfahan, Iran, h.bagheri@cet.ui.ac.ir}


		\begin{abstract}
			\textcolor{blue}{This is the pre-acceptance version, to read the final version, please go to Urban Climate on ScienceDirect, \url{https://www.sciencedirect.com/science/article/abs/pii/S2212095524002992}}. Climate models lack the necessary resolution for urban climate studies, requiring computationally intensive processes to estimate high resolution air temperatures. In contrast, Data-driven approaches offer faster and more accurate air temperature downscaling. This study presents a data-driven framework for downscaling air temperature using publicly available outputs from urban climate models, specifically datasets generated by UrbClim. The proposed framework utilized morphological features extracted from LiDAR data. To extract urban morphological features, first a three-dimensional building model was created using LiDAR data and deep learning models. Then, these features were integrated with meteorological parameters such as wind, humidity, etc., to downscale air temperature using machine learning algorithms. The results demonstrated that the developed framework effectively extracted urban morphological features from LiDAR data. Deep learning algorithms played a crucial role in generating three-dimensional models for extracting the aforementioned features. Also, the evaluation of air temperature downscaling results using various machine learning models indicated that the LightGBM model had the best performance with an RMSE of 0.352\textdegree K and MAE of 0.215\textdegree K. Furthermore, the examination of final air temperature maps derived from downscaling showed that the developed framework successfully estimated air temperatures at higher resolutions, enabling the identification of local air temperature patterns at street level. The corresponding source codes are available on GitHub: \url{https://github.com/FatemehCh97/Air-Temperature-Downscaling}.
		\end{abstract}
		
	\end{frontmatter}
	
	\section{Introduction}\label{sec.intro}
	With the growing trend of urbanization worldwide, urban areas are undergoing rapid development, leading to the emergence of unique climatic conditions known as urban climate \citep{RN135, RN138}. Urban climate is a complex phenomenon formed by a combination of factors such as regional climate patterns, human activities, industrial and commercial activities, land-use patterns, etc. \citep{RN137, RN136, zhu2020so2sat, ganjirad2024google}. As urbanization continues, changes in urban climate are increasing globally. Urban climate change is one of the challenges of the city, which has extensive impacts on health, livelihoods, economy, infrastructure, services, and ecosystems \citep{RN143, RN67}. Therefore, understanding urban climate is essential for efficient urban planning, sustainable development, and ensuring the well-being of urban residents \citep{RN139}. Urban climate models enable the simulation, prediction, and assessment of various climate scenarios by analyzing different meteorological variables, providing valuable insights for urban planners and decision-makers \citep{RN77}.
	
	Research on climate modeling is constantly evolving thanks to the development of new sensors and measurement techniques, increased computational power, and advanced knowledge. Among the various climate parameters, air temperature is one of the key variables of urban climate. Several models have been designed for urban climate modeling, including machine learning and deep learning-based models \citep{RN90}, GIS (Geographic Information System)-based models such as UMEP (Urban Multi-scale Environmental Predictor) \citep{RN75}, and physics-based models like UrbClim (Urban Climate modeling) \citep{RN109}, ENVI-met \citep{RN95}, WRF (Weather Research and Forecasting Model) \citep{RN92}, and UWG (Urban Weather Generator) \citep{RN93}. Despite the development of various models, there are still many challenges in the field of climate modeling for urban areas. One of these challenges is the mismatch between the scale of climate data suitable for urban planners and designers and the scale of climate data provided by climate models. Urban planners and designers typically require medium to fine-scale information about urban systems, while the outputs of climate models are generally too coarse and do not capture the details and complexity of urban systems \citep{RN209}. Therefore, various techniques have been developed for downscaling the outputs of climate models.
	
	One of the well-known physical  model used for climate modeling in urban area is UrbClim.  This model is designed to study and simulate urban climate variables such as air temperature, wind speed, and humidity at the local scale with a spatial resolution of 100 m. This model reduces large-scale weather conditions to the agglomeration-scale. Compared to other mesoscale models, UrbClim has a higher execution speed and requires less computational power. The UrbClim model has been validated for several European cities and has provided satisfactory results \citep{RN109}. While this model has been developed for estimating air temperature at the urban scale, taking into account urban morphology and land use, it has limitations in terms of spatial resolution. For many urban applications, such as modeling air temperature variations at the building level, studying the impact of tree shading on urban comfort, etc., higher resolutions estimation of (e.g., 5 m) of air temperature are required \citep{RN234}.
	
	One of the challenges in air temperature estimation at very high resolutions is the computational limitation of physically-based downscaling methods, which require a high computational cost for air temperature estimation at higher resolutions. To address this issue, statistical downscaling methods using machine learning models can be utilized. Although previous studies have employed downscaling approaches using ground-based data to achieve higher resolution air temperature estimation \citep{RN10, RN203}, directly applying machine learning methods for downscaling air temperature at the city level requires a large number of observation stations with appropriate distribution throughout the city, which generally requires a significant cost. On the other hand, machine learning models are not capable of accurately modeling and predicting events with extreme variations, such as sudden heat or cold waves. One advantage of downscaling the outputs of physical models using machine learning is the ability to predict abrupt temperature changes at a larger scale. Thus, at lower resolutions (e.g., 100 m), physically-based models like UrbClim can estimate air temperature and its abrupt variations at the city level, and then machine learning models can be employed for downscaling of the outputs of physical models to obtain much higher resolution with lower computational costs.
	
	The main objective of this research is to develop and implement a machine learning-based framework for downscaling the air temperature estimates derived from physical urban climate models such as UrbClim to achieve higher resolutions. Considering that air temperature at microclimate scale is influenced by urban morphology \citep{RN108, RN198, RN192, RN199, RN200}, urban morphological features are extracted from LiDAR data in the proposed framework. Another contribution of this study is to investigate how to extract urban morphological parameters from LiDAR data. The ultimate goal is to evaluate the proposed framework and the potential of LiDAR data in downscaling the UrbClim model to obtain maximum, minimum, and average daily air temperature maps with a resolution of 5 m. Furthermore, this research paves the way for developing a digital twin of the urban climate model.
	
	This paper consists of several sections. In current section, we outlined the main objectives of the paper. A review of previous investigations is provided in Section \ref{sec_lit}. The specifications of the study area and the data used in this research are presented in Sections \ref{sec_2} and \ref{sec3_data}, respectively. In Section \ref{sec_4}, the proposed framework, which includes data preparation, extraction of building footprints, generation of the 3D building model, extraction of morphological features from the 3D model, and the development of a machine learning model for downscaling air temperature, is described. The results obtained from various experiments as well as  the evaluation of the proposed framework's capability in air temperature downscaling for achieving higher resolutions are presented and discussed in Sections \ref{sec_5} and \ref{sec_6}, respectively.
	
	\section{\textcolor{red}{Literature Review}}\label{sec_lit}
	While urban climate models like UrbClim provide accurate simulations at a resolution of 100 m, many applications in urban management and environmental areas require urban microclimate data at a higher level of detail. In urban planning, higher resolution modeling of urban climate allows municipalities and urban planners to make better decisions regarding urban development, design, and green space management \citep{CORTES202297, atmos10050282}. The ability to accurately identify air pollution hotspots and their environmental impacts, determine the best methods for managing heat islands and air pollution \citep{RN246}, and cope with natural disasters such as heatwaves \citep{huttner2008using} or floods \citep{su11226361} all require high resolution urban climate data.
		Various investigations have focused on downscaling the urban climate data to achieve higher resolutions, such as 10 m \citep{DADIOTI2017366, gmd-15-5309-2022, gmd-13-1335-2020}.
	Different studies have categorized downscaling techniques into three groups: dynamical, statistical, and hybrid approaches \citep{RN209, RN210,RN211, RN205}. 
	
	Dynamical downscaling methods utilize Regional Climate Models (RCMs) or numerical atmospheric models to generate outputs with higher resolutions. For example, \citeauthor{RN208} (\citeyear{RN208}) focused on investigating and predicting urban heat islands (UHI) in the Brussels region of Belgium. In this study, three dynamical downscaling methods were compared to improve the accuracy of simulations at the urban scale (with a resolution of 1 km). The study showed that the three downscaling methods could successfully simulate the UHI effect. \citep{RN208}. Dynamical downscaling methods have a high capability for simulating, estimating, and predicting urban weather, particularly extreme events. However, these models require significant computational power due to the use of numerical methods in the downscaling process, which makes their implementation challenging for high spatial resolutions (at the microclimate scale) \citep{RN209}. 
	
	In statistical downscaling methods, statistical relationships between large- and small-scale climate variables are employed to transform model outputs into higher resolutions. Various statistical methods have been developed to enhance the resolution of climate data, including machine learning methods \citep{RN204}, regression approaches, analog methods, and weather typing techniques \citep{RN213}.
	
	\textcolor{red}{\citeauthor{RN213} (\citeyear{RN213}) evaluated 12 statistical downscaling methods for air temperatures in Spain, finding regression methods most suitable for climate change studies due to their strong correlation with observed data \citep{RN213}.}
	
	As mentioned earlier, due to the complexity and high computational costs of physical models in achieving extremely higher resolutions for urban air temperature estimation, using machine learning methods to achieve higher resolution air temperature estimation may be more efficient. In various studies, different machine learning methods, e.g., multi-variable linear regression, have been employed to downscale air temperature using various data sources such as air temperature networks, land cover information, remote sensing imagery, and crowdsourced weather data. These studies achieved resolutions ranging from 10 to 30 m, depending on the resolution of input data  \citep{CHEN2019710, pnas.1817561116, VENTER2020111791, YIN2020101538}. \citeauthor{DING2023110211} (\citeyear{DING2023110211}) recently developed a model by incorporating social-economic information and downscaled land cover data to achieve hyper-local air temperature mapping (1m resolution) \citep{DING2023110211}.
	
	\textcolor{red}{\citeauthor{RN225} (\citeyear{RN225}) compared  Multiple Linear Regression (MLR), Artificial Neural Network (ANN), and Least Squares Support Vector Machine (LS-SVM) for monthly minimum and maximum air temperature prediction in central India's Tons River basin, finding LS-SVM to be the most accurate \citep{RN225}.}
	
	\textcolor{red}{\citeauthor{RN226} (\citeyear{RN226}) evaluated the Random Forest (RF) for downscaling air temperature in southern China's Pearl River basin, showing it outperformed ANN, MLR, and SVM in accuracy and robustness \citep{RN226}.}
	
	\textcolor{red}{\citeauthor{RN204} (\citeyear{RN204}) developed the Building-Scale Resolved Air Temperature (BRT) model for estimating air temperature at the building and street level in Seoul, South Korea, improving accuracy with a 25 m resolution. The study found that SVM outperformed linear regression \citep{RN204}.}
	
	\textcolor{red}{\citeauthor{RN223} (\citeyear{RN223}) compared Long Short-Term Memory (LSTM), SVM, MLR, and arithmetic ensemble mean for long-term daily air temperature downscaling in Ontario, Canada. Both statistical and machine learning methods showed high accuracy but struggled with extreme temperatures. Mean ensembles were recommended for lower computational requirements \citep{RN223}.}
	
	\textcolor{red}{\citeauthor{RN10} (\citeyear{RN10}) used machine learning to study the correlation between urban morphology and air temperature in Singapore, finding that machine learning models, particularly the voting regression algorithm, provided more accurate predictions than conventional linear models \citep{RN10}.}
	
	\textcolor{red}{\citeauthor{RN203} (\citeyear{RN203}) downscaled urban nighttime air temperatures using weather stations and satellite data in four major Lebanese cities, reducing spatial resolution from 25 km to 1 km \citep{RN203}.}
	
	\textcolor{red}{\citeauthor{RN224} (\citeyear{RN224}) compared RCMs and statistical downscaling models (SDMs) for weather variable downscaling across Europe, finding SDMs generally outperformed RCMs. After bias correction, both models produced similar outcomes, highlighting SDMs' potential in downscaling global climate predictions \citep{RN224}.}

	Most studies have realized air temperature estimation at higher resolutions based on dense ground stations or crowdsourced weather measurements. However, crowdsourced data may not always be as accurate or reliable as data collected by professional meteorologists and weather stations. On the other hand, the installation and operation of many ground stations demand high deployment, establishment, and maintenance costs. This study uses estimates from physical models such as UrbClim at lower resolution as a reference and proposes a downscaling procedure to provide higher resolution air temperature estimates.
	
	\section{Study Area}\label{sec_2}
\begin{figure}[!t]
		\centering
		\captionsetup{justification=centering}
		{\includegraphics[width=\linewidth]{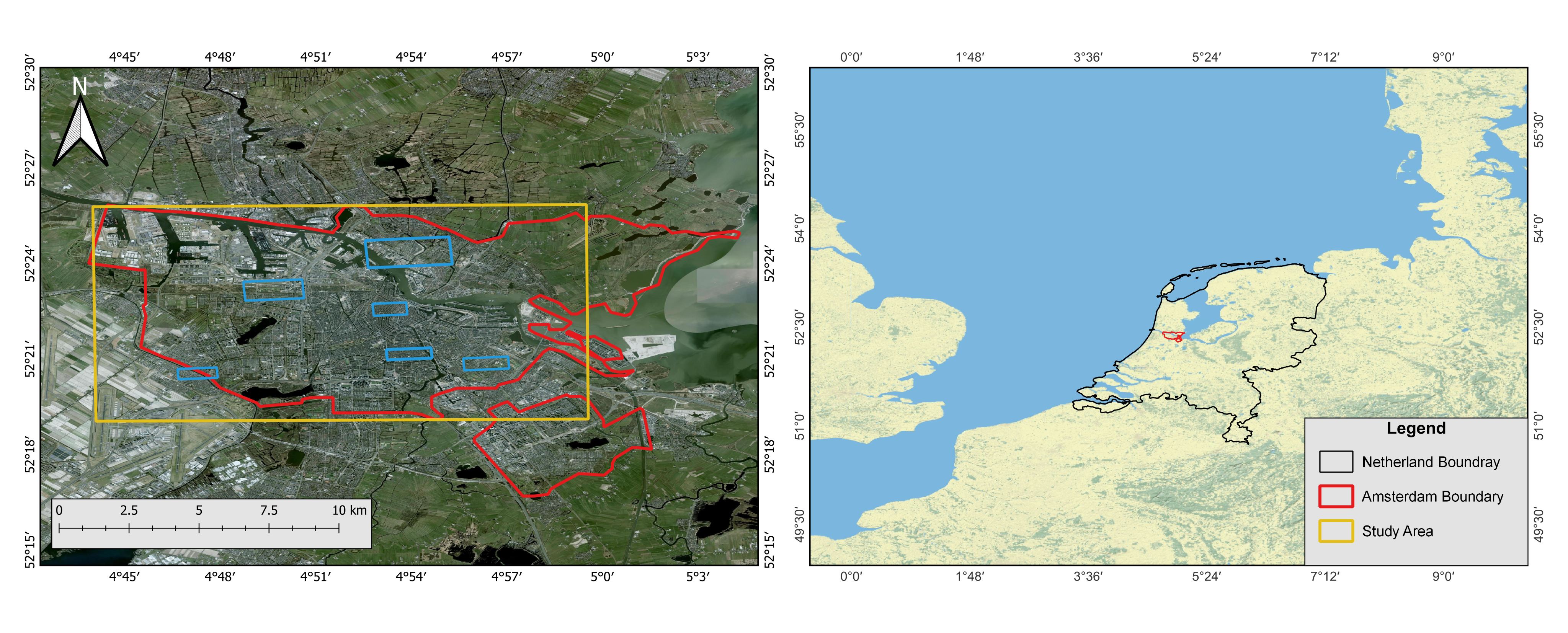}}
		\vspace{0.05cm}
		\caption{A visualization of the study area, Amsterdam, the capital of the Netherlands. The red boundary represents the city limits, the yellow rectangle represents the study area, and the blue rectangles represent the training areas.}
		\label{study_area}
	\end{figure}

	This study was conducted in Amsterdam, the capital of the Netherlands. The city is located in the western part of the Netherlands, adjacent to the Amstel River. Amsterdam has a total area of 219.4 square km, with 12\% of it consisting of green spaces \citep{RN134}. The city is situated 2 m below sea level \citep{RN215}. Amsterdam has an oceanic climate, generally mild and humid, influenced by its proximity to the North Sea and the winds that rise from it \citep{RN217}. Amsterdam has a high diversity in land use, encompassing residential, commercial, industrial, and green areas \citep{RN180}. Moreover, the buildings vary in height and density. This diversity in building structure (morphology) and land use makes Amsterdam a challenging, valuable study area for urban climate research. Fig.\ref{study_area} illustrates the extent of Amsterdam city and the study area, which covers an area of 219.32 square km.
	
	\section{Datasets}\label{sec3_data}
	\subsection{Urban Climate Data}
	\textcolor{red}{In this study, we utilized meteorological data, including air temperature, wind speed, and relative humidity with a spatial resolution of 100 m and hourly temporal resolution, obtained from pre-existing datasets generated by the UrbClim physical model and provided by the Copernicus Climate Change Service \href{https://cds.climate.copernicus.eu/}{(https://cds.climate.copernicus.eu/)}.} The UrbClim model has been designed to simulate and study urban climate variables such as the urban heat island effect, wind speed, and air humidity. This model relies on two types of primary input data: large-scale meteorological data and the description of the urban terrain (terrain-related inputs including land use, soil sealing, vegetation coverage, topography, and anthropogenic heat flux). The model was initially developed at the Flemish Institute for Technological Research (VITO) \citep{RN109, RN240}. Then, \citeauthor{RN181} further developed the UrbClim model using ERA5 data and obtained climate variables for one hundred European cities from 2008 to 2017. Based on validation assessments conducted in various European cities, it was concluded that this model exhibited sufficient accuracy ($R^{2} >= 0.96$) in urban air temperature estimation \citep{RN109}. Consequently, its outputs can be considered a reliable baseline for this study. The UrbClim outputs for January, May, July, and October 2017 were used in this study.

	\subsection{LiDAR Data}
	The LiDAR point cloud data for Amsterdam are part of the Dutch elevation data (AHN3) provided by Actueel Hoogtebestand Nederland \href{https://ahn.nl/}{(AHN)}. These data are available in the LAZ format, a compressed version of the LAS file format \href{https://app.pdok.nl/ahn3-downloadpage/}{(https://app.pdok.nl/)}. The AHN3 data for the Netherlands were collected between 2014 and 2019, with a vertical accuracy of less than 5 cm and an average point density ranging from 6 to 10 points per square meters \citep{RN215}.
	
	In addition to the LiDAR data of Amsterdam, LiDAR data for some parts of Miami-Dade County, Florida, USA, in the form of a normalized Digital Surface Model (nDSM) with a resolution of one foot (0.3 m), were also utilized in this study. As will be explained in Section \ref{sec_4.2.2}, these data are used for the initial training of semantic segmentation models for building footprint detection \href{https://coast.noaa.gov/dataviewer/#/lidar/search/where:ID=9041}{(https://coast.noaa.gov/)}.
	
	\subsection{OpenStreetMap Data}
	OpenStreetMap (OSM) data of Amsterdam has been utilized to extract urban morphological features and prepare training data for deep learning-based footprint extraction. For this purpose, three layers of OSM data have been used, including the Landuse layer (park and forest classes), the Water layer, and the Buildings layer \href{https://www.openstreetmap.org/}{(https://www.openstreetmap.org/)}. The latest update of these datasets is for 2022. The Landuse and Water layers are used to create features such as distance to water, distance to parks, and pervious surfaces. The Buildings layer is also employed for transfer learning techniques in building footprint detection.
	
	\section{Methods}\label{sec_4}
\begin{figure}[!t]
		\centering
		{\includegraphics[width=1\linewidth]{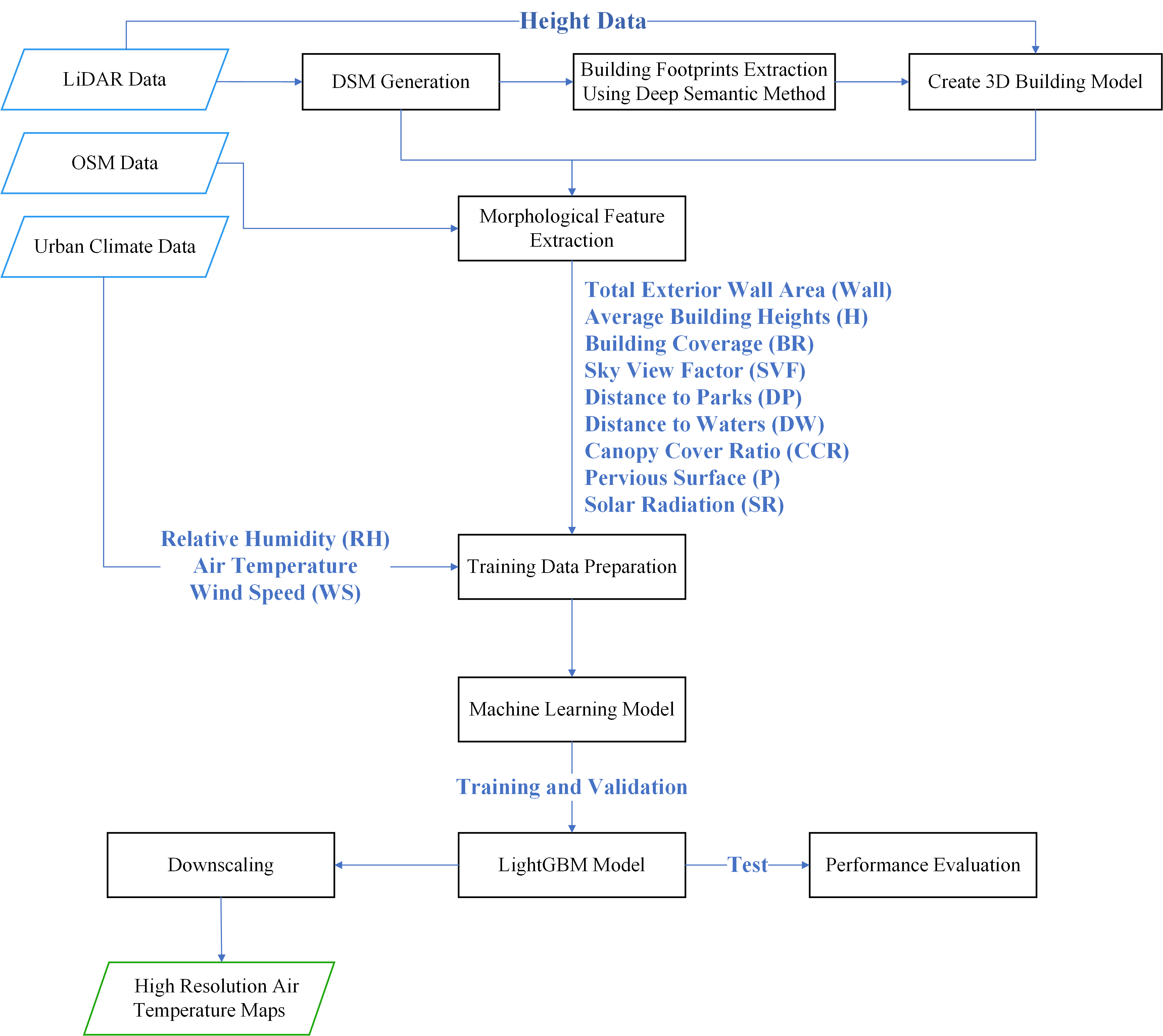}}
		\vspace{0.05cm}
		\caption{The framework designed for downscaling of air temperature over the study area}
		\label{methodology_flowchart}
	\end{figure}

	To achieve higher resolution in estimating urban air temperatures while considering the influence of urban morphology and buildings, a framework based on machine learning techniques has been proposed and implemented in this study. In this framework, air temperatures obtained from the urban climate physical model, UrbClim, with lower resolution (100 m) are downscaled to higher resolution (5 m). Fig.\ref{methodology_flowchart} illustrates the different stages of the proposed framework. The framework consists of five key stages, including data preparation, extraction of building footprints using deep learning techniques, creation of a 3D building model, extraction of parameters used for air temperature downscaling, and finally, the development of a machine learning model for microclimate temperature estimation. As mentioned in Section \ref{sec3_data}, LiDAR data, urban climate data, and OSM data are used to achieve this objective. Firstly, a Digital Surface Model (DSM) is generated from the LiDAR point cloud data. Then, deep learning models are employed to extract building footprints from the DSM. Using elevation information from LiDAR point clouds and the extracted footprints, a three-dimensional model of buildings is created in the study area. Next, urban morphological features are derived from the generated 3D model and some layers of OSM data. In addition to these parameters, daily weather data are prepared. Subsequently, machine learning models are trained using the extracted morphological and meteorological parameters, and air temperature data from the UrbClim model. Finally, the trained machine learning model is used to downscale air temperatures obtained from the UrbClim model from a resolution of 100 m to 5 m for the urban area of Amsterdam. Each of these stages is described in more detail in the following sections.

	\subsection{Data Preparation}
	The data preparation process includes converting LiDAR point cloud data into DSM and preparing urban climate data derived from the UrbClim model. The LiDAR data initially consists of three-dimensional XYZ coordinates. To convert this three-dimensional data into a DSM, the "Adaptive Triangulation" method is used. The resulting DSM has a resolution of 0.23 m, providing a detailed representation of the ground surface and buildings. The created DSM for the study area is shown in Figure S1 in the supplementary.
	The urban climate data includes air temperature, wind speed, and relative humidity in netCDF4 format, available as hourly data for each month. Initially, the mentioned data are extracted, and the average, minimum, and maximum values of air temperature, wind speed, and relative humidity are calculated for each day in the selected four months, January, May, July, and October. Additionally, for each day, the wind speed and relative humidity values corresponding to the minimum and maximum air temperatures are extracted. In more detail, for each day, the hour at which the air temperature reaches its minimum and maximum is identified and stored as the minimum and maximum air temperature, respectively. The wind speed and relative humidity values corresponding to these hours are selected as the minimum and maximum values for that day to obtain the wind speed and relative humidity values associated with the coldest and warmest air temperatures.
	
	\subsection{Building Footprints Extraction}\label{sec_4.2}
	As shown in the framework in Fig.\ref{methodology_flowchart}, the first step for downscaling air temperature is the extraction of building footprints to create a 3D model. This 3D model is then used to extract the required features. Several studies have been conducted to extract buildings from remote sensing data \citep{RN175,RN4,RN220,RN42}. In most of these studies, buildings were extracted from high resolution aerial or satellite images. However, due to the unavailability of such data for the study area, in this research, the possibility of extracting building footprints from LiDAR data was explored. Extracting buildings from LiDAR data is a more challenging process compared to aerial and satellite images, mainly due to the lack of different spectral bands. To achieve this goal, image segmentation models using deep learning algorithms were employed.
	
	\subsubsection{\textcolor{red}{Deep Semantic Segmentation}}

	Deep learning-based semantic segmentation is a computer vision task that is used to classify each pixel in an image into a particular class \citep{RN187}. This study employed semantic segmentation to detect building footprints from DSM data derived from LiDAR, with building and non-building classes as label images. During training, the model learns to distinguish building pixels from non-building pixels and performs semantic segmentation for the DSM data (identifying the buildings). In this study, various deep learning models were employed for building footprint detection, including U-Net and its variants (Attention U-Net and U-Net3+), as well as DeepLabV3+.

	The U-Net architecture is one of the most popular image segmentation models. This architecture is based on convolutional neural networks (CNNs), initially developed for medical image segmentation \citep{RN110}. The network consists of an encoder-decoder structure with skip connections that preserve spatial information and improve segmentation accuracy. The encoder extracts contextual information from the input image while the decoder reconstructs the segmented image (building and not-building pixels). Attention U-Net, a variant of the U-Net model, leverages attention mechanisms within its skip connections to enhance segmentation performance \citep{RN112}, by focusing on relevant image features and reducing the importance of irrelevant ones \citep{RN126}. The inclusion of attention gates in the skip connections significantly enhances the segmentation performance without adding excessive computational complexity. U-Net3+ is another improved segmentation structure upon the original U-Net by employing full-scale skip connections and deep supervision \citep{RN111}. The design of connections in this model allows both low resolution and high resolution feature maps to be utilized for segmentation. The model further benefits from a hybrid loss function and a classification-guided module (CGM), enhancing segmentation accuracy. This model improves accuracy and reduces the number of network parameters \citep{RN111}. Another model employed is DeepLabV3+, the latest version of the DeepLab semantic segmentation model developed by Google AI \citep{RN148}. This model combines the spatial pyramid pooling (SPP) module with an encoder-decoder structure \citep{RN149, RN151}. It uses atrous convolutions in the SPP to control feature resolution and balance accuracy and execution time \citep{RN152, RN214}. This Atrous Spatial Pyramid Pooling (ASPP) module corrects feature maps using multi-scale information.

	\subsubsection{Transfer Learning}\label{sec_4.2.2}
\begin{figure}[!t]
		\centering
		{\includegraphics[width=\linewidth]{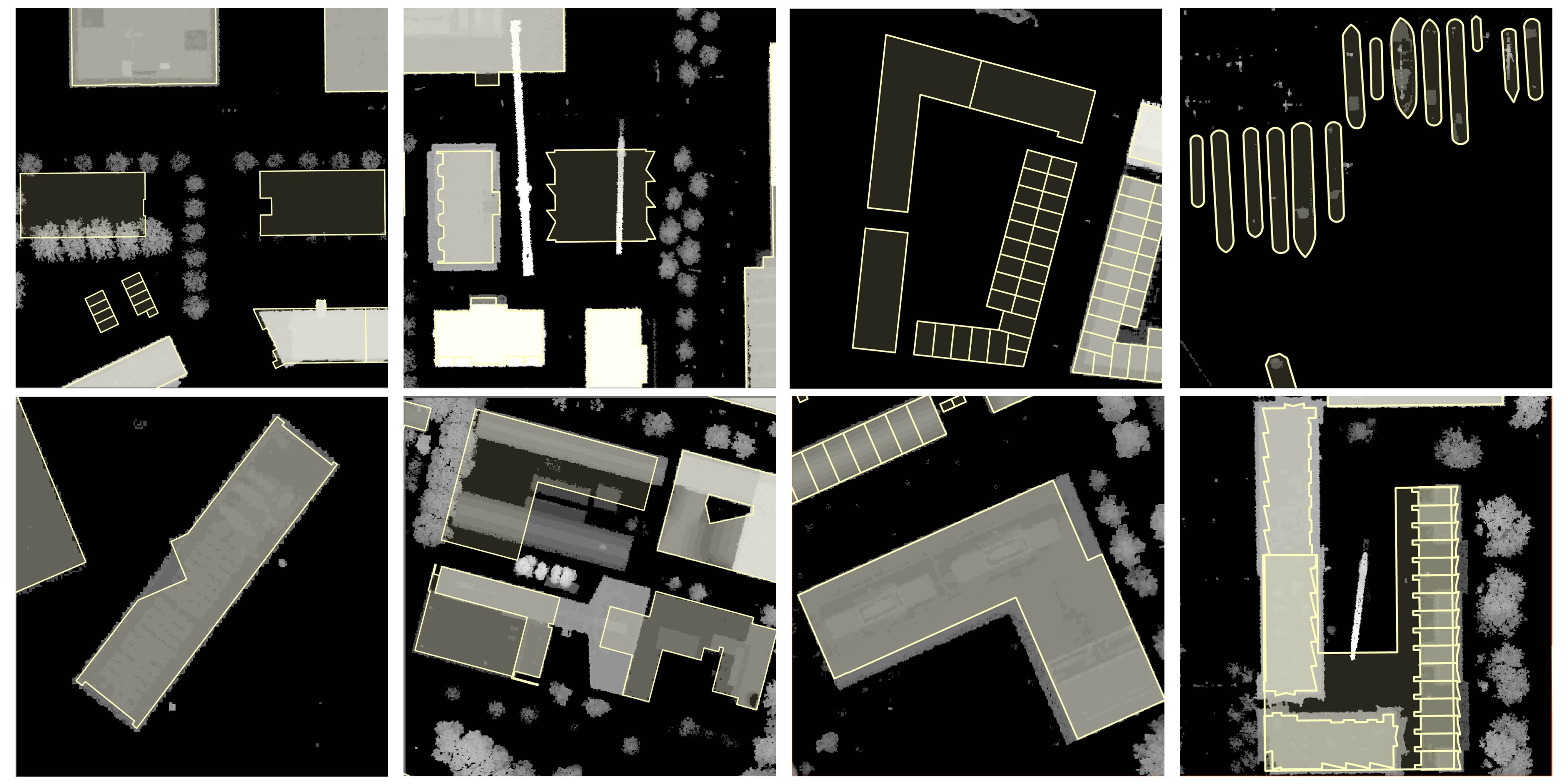}}
		\vspace{0.05cm}
		\caption{An example of samples to show discrepancies between OSM building Layer and LiDAR ground truth. OSM building footprints are depicted as yellow polygons and LiDAR DSM is visualized as a base map.}
		\label{osm_fig}
	\end{figure}

	Transfer learning is a technique that utilizes the knowledge acquired by a pre-trained model to solve a different but related task \citep{RN129}. In the context of this study, the footprint data of existing buildings in OSM in the study area may not necessarily be complete or accurately positioned \citep{RN190,RN147,RN189}. Therefore, it is necessary to accurately extract these footprints from LiDAR data. On the other hand, due to limited access to diverse training data (labels), training the mentioned deep learning models solely relying on limited Amsterdam data is not feasible to achieve satisfactory and acceptable performance. Consequently, transfer learning can be used as a solution. In this study, deep learning models were initially trained on LiDAR data from the city of Miami-Dade, which had accurate labels. Then, transfer learning was employed to enhance the performance of the pre-trained model using Miami-Dade data for detecting building footprints in the Amsterdam region. For this purpose, the model with the best performance on Miami-Dade data was selected and further trained with Amsterdam data. LiDAR data along with the building footprint layer provided by OSM were used as the training data for Amsterdam. Fine-tuning the model with Amsterdam data allows the model to better generalize to the structural differences between buildings in Miami-Dade and Amsterdam, resulting in improved performance in the new geographical area. However, the provided building footprints by OSM in different areas of Amsterdam also have multiple defects, and it is not possible to fully rely on them for building modeling. Fig.\ref{osm_fig} illustrates some examples of defects in the OSM building layer, including discrepancies between the footprints and the actual shapes of some buildings. There are buildings in recent years that have not been captured in the LiDAR data due to the time gap between the data collection, and there are cases in OSM where boats have been mistakenly identified as buildings.
	
	\subsubsection{Model Setup and Performance Evaluation}
In Table \ref{table_DLsetup}, the architectures and training details of each segmentation model are presented. This table summarizes the hyperparameters tuned for each structure, including the type of pooling, activation functions, upsampling method, optimizer, type of loss, and batch size. Additionally, the structural details of the algorithms, such as the employed layers and the number of filters in the model architectures, are depicted in the figures presented in Appendix \ref{appendix_dl_fig}. These parameters and model structures were designed to minimize the validation loss and improve overall performance. Furthermore, data augmentation techniques were employed to increase the training dataset and enhance the model's generalization ability. These techniques included rotation, width and height shifting, horizontal flipping, zooming, and shearing. All models were trained using an NVIDIA-RTX 3090.

In the U-Net3+ model, a feature aggregation mechanism was applied to the concatenate feature map of five different scales to integrate fine-grained details from shallow layers with high-level semantic information. This mechanism included Conv2D with 160 filters of size 3×3, Batch Normalization, and ReLU activation.

In the DeepLabV3+ model, a pre-trained ResNet50 model trained on ImageNet was used as the backbone. The input to ResNet50 is three-channel images; however, since the input images in this study are single-channel, they are initially transformed into three-channel images. In the encoder section, the ASPP technique with four dilation rates of 1 (regular convolution), 6, 12, and 18 was applied. Then, the features were passed through a 1x1 convolution layer and upsampled using bilinear interpolation with a factor of 4. Finally, they were concatenated with the low-level feature map with the same spatial resolution from the backbone network in the decoder.

There are various metrics for evaluating segmentation results, which are generally pixel-based. In this type of evaluation, the predicted building footprints are compared to the reference footprints for each pixel in the test images. Segmentation results were evaluated using IoU (Intersection over Union), accuracy, precision, recall, and f1-score metrics. These metrics provide insights into the models' effectiveness in accurately identifying building footprints.
	
	\begin{table}[t]
		\centering
        \caption{Deep learning models architectures and training details for building footprint extraction}
		\label{table_DLsetup}
		\begin{adjustbox}{width=\textwidth}
			\begin{tabular}{c|c|c|c|c}
				\diagbox[width=10em]{Setup}{Models} & U-Net & Attention U-Net & U-Net3+ & DeepLabV3+ \\
				\toprule
				\makecell{Input Dimensions \\ (H x W x D)} & (512, 512, 1) & (512, 512, 1) & (512, 512, 1) & (512, 512, 3) \\
				\midrule
				Number of Filters & [64, 128, 256, 512, 1024] & [64, 128, 256, 512, 1024] & [64, 128, 256, 512, 1024] & Backbone = ResNet50 \\
				\midrule
				\makecell{Activation \\ Functions} & \makecell{ReLU (Hidden) \\ Sigmoid (Output)} & \makecell{ReLU (Hidden) \\ Sigmoid (Output)} & \makecell{ReLU (Hidden) \\ Sigmoid (Output)} & \makecell{ReLU (Hidden) \\ Sigmoid (Output)} \\
				\midrule
				\makecell{Downsampling \\ Method} & \makecell{MaxPooling2D \\ (2x2, stride 2)} & \makecell{Strided Conv2D + \\ batch norm + \\ activation} & \makecell{Strided Conv2D + \\ batch norm + \\ activation} & ASPP \\
				\midrule
				\makecell{Upsampling \\ Method} & \makecell{Conv2DTranspose \\ + batch norm \\ + activation} & \makecell{Conv2DTranspose \\ + batch norm \\ + activation} & \makecell{Conv2DTranspose \\ + batch norm \\ + activation} & \makecell{2D UpSampling \\ (bilinear)} \\
				\midrule
				Optimizer & Adam (LR: $10^{-4}$) & Adam (LR: $10^{-4}$) & Adam (LR: $10^{-4}$) & Adam (LR: $10^{-4}$) \\
				\midrule
				Loss & Binary Cross Entropy & Binary Cross Entropy & Hybrid Loss & Binary Cross Entropy \\
				\midrule
				Batch Size & 8 & 8 & 4 & 8 \\
				\midrule
				Structure & Fig.\ref{unet_fig} & Fig.\ref{attention_fig} & Fig.\ref{unet3p_fig} & Fig.\ref{deeplab_fig} \\
			\end{tabular}
		\end{adjustbox}
	\end{table}

	\subsubsection{Postprocessing of Detected Building Footprints}\label{sec_4.2.5}
\begin{figure}[!t]
		\centering
		{\includegraphics[width=0.55\linewidth]{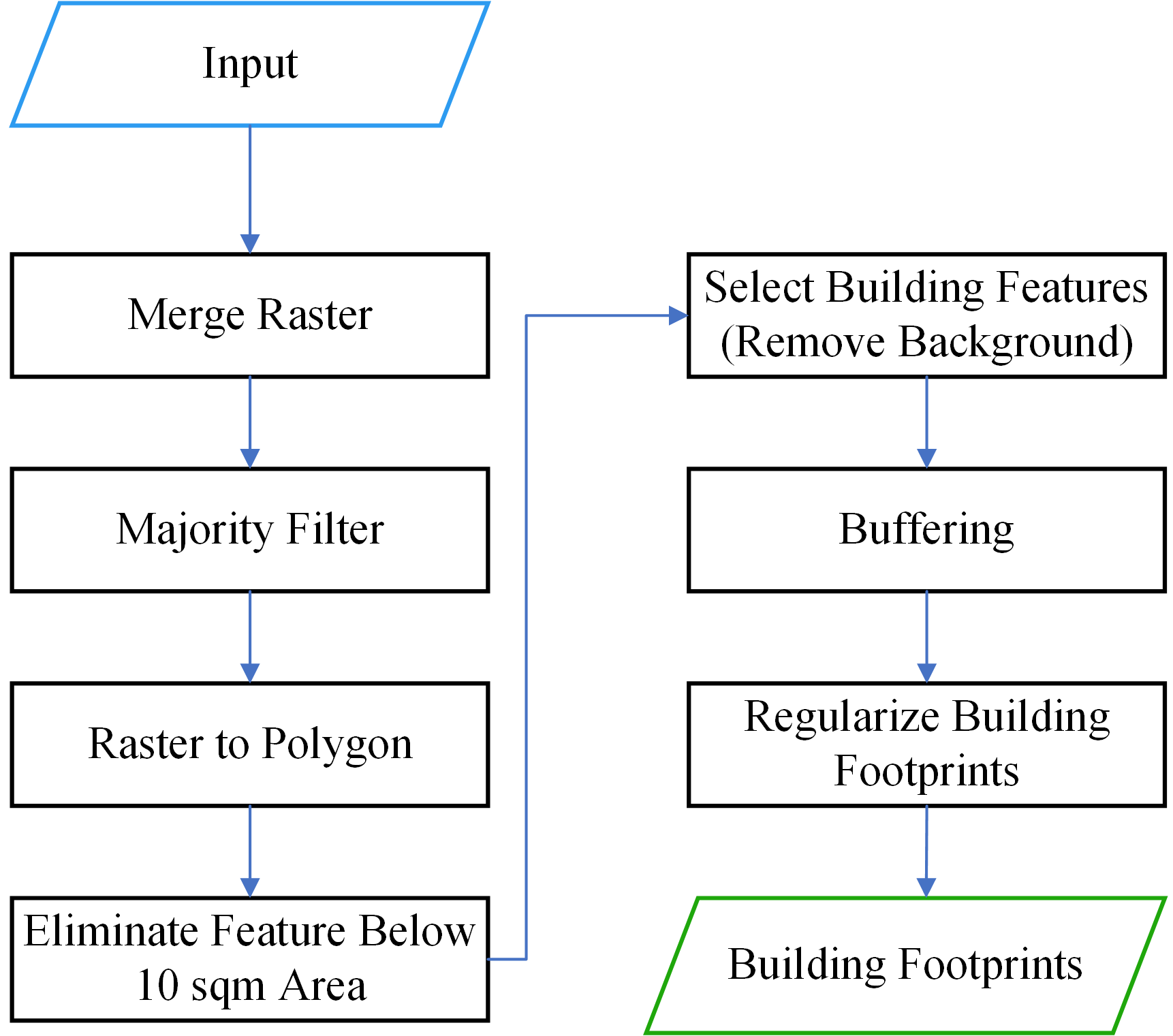}}
		\vspace{0.05cm}
		\caption{Postprocessing on the output of segmentation to improve the quality of building footprint extraction}
		\label{postprocess_flowchart}
	\end{figure}

	The model's output is in the form of binary rasters of size 512 $\times$ 512, where pixels corresponding to buildings are assigned a value of 1 and other pixels are assigned a value of 0. This output may not be perfect, so various post-processing steps are performed to improve the model's output, correct possible errors, and ultimately generate vectorized building footprints. Fig.\ref{postprocess_flowchart} illustrates the post-processing steps performed to achieve the final building footprints. According to Fig.\ref{postprocess_flowchart}, in the first step, the model outputs of size 512 $\times$ 512 were merged to obtain a single integrated raster of predictions. In the next step, a 3$\times$3 Majority Filter was applied to the integrated raster to reduce noise. This filter replaces each cell in the raster with the majority of its neighboring cells (8 neighboring cells). Then, the building segments in the raster were converted to polygons and vectorized. In the next operation, small areas that were detected as buildings were eliminated. For this purpose, features that were smaller than a certain threshold (10 square meters) were selected and eliminated. After selecting the building features, and removing the background, the boundaries of the buildings were slightly increased by creating a buffer for the polygons to a specified distance (5 cm). Then, the building polygons were normalized by removing undesirable geometric flaws, and the final footprints of the buildings were generated. This was realized by employing Regularize Building Footprint (RBF) process. The RBF process utilizes the polyline compression algorithm to correct undesirable distortions in the building footprints \citep{RN228}. Finally, any remaining issues resulting from this process were manually addressed.
	
	\subsection{Generating 3D Buildings Model}\label{sec_4.3}
	3D city modeling is performed at various levels of detail. The concept of level of detail (LOD) is used to define a set of different representations of real-world objects and express the complexity of the models, which has a significant impact on the usability and applicability of the models \citep{RN130, RN237, bagheri2019fusion}. Displaying different levels of detail according to the diverse needs of users leads to reducing complexity, computational cost, and optimal utilization of processing power. CityGML, as an Open Geospatial Consortium (OGC) standard for representing and storing 3D city models, defines five levels of detail (LODs 0-4) \citep{RN87}. 
	Our study focused on generating LOD1 models, representing buildings as simple blocks with flat roofs. This level of detail provides a basic three-dimensional representation of buildings, enhancing visual understanding while minimizing computational complexity.
	
		%
	
\begin{figure}[!t]
		\centering
		{\includegraphics[width=0.55\linewidth]{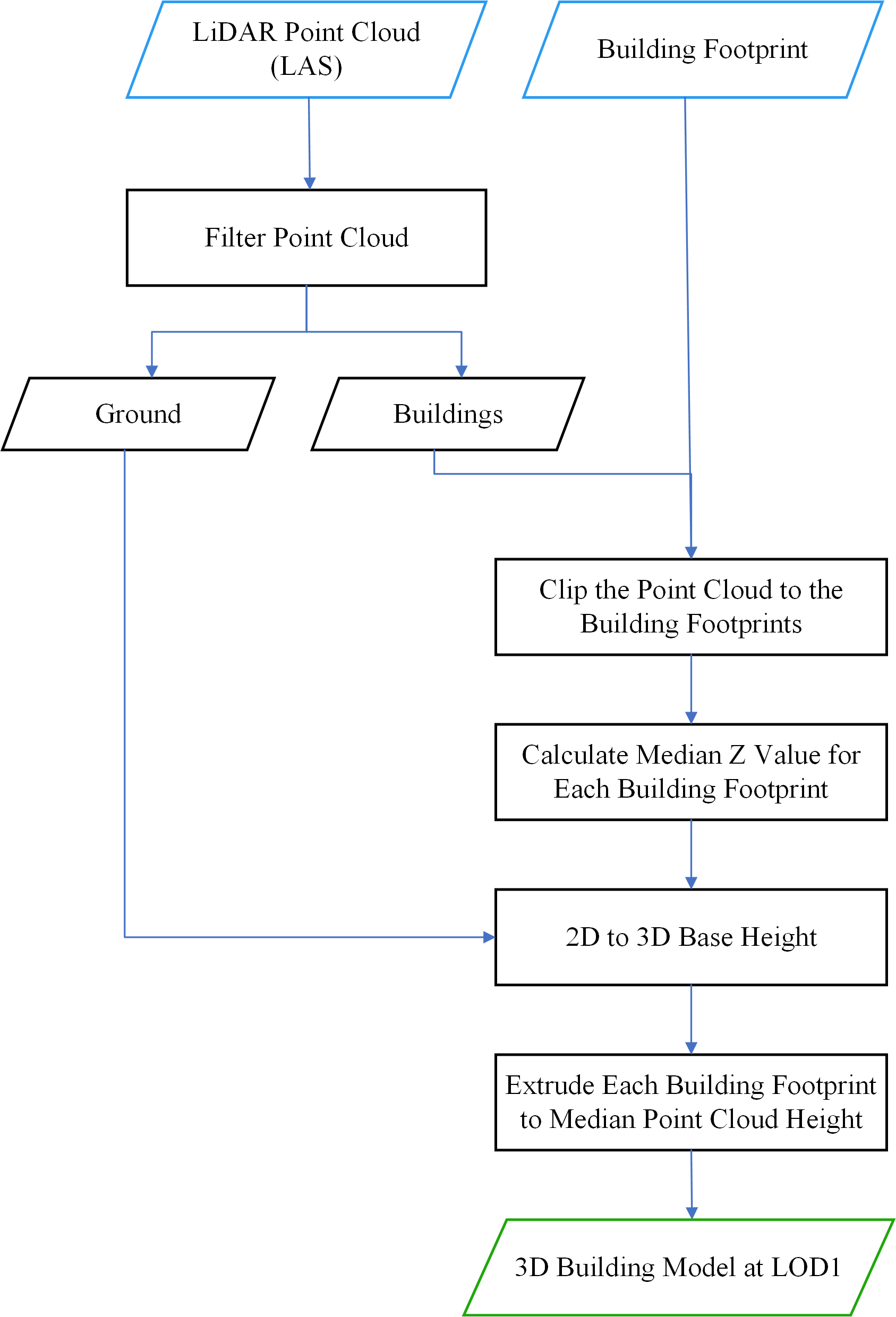}}
		\vspace{0.05cm}
		\caption{3D city model generation at LOD1 using LiDAR point clouds and building footprints}
		\label{3d_flowchart}
	\end{figure}

	After detecting the footprints of the buildings, a 3D model of the buildings at LOD1 was created using the height information obtained from airborne LiDAR point clouds. This process was carried out in accordance with the flowchart shown in Fig.\ref{3d_flowchart}. In the first step, ground and non-ground points in the LiDAR point cloud were separated by filtering the point cloud based on predefined labels. The ground points are used to estimate the base height and generate a Digital Terrain Model (DTM). Next, based on the building footprints, the LiDAR point cloud was clipped to retain only the points related to the buildings. Then, the median height value was calculated for each building. In the next step, the 2D building footprints were transformed into 3D space using the base height (DTM) obtained from the LiDAR point cloud. Finally, each building footprint was extruded based on the median height derived from the LiDAR point cloud. This process resulted in the generation of a 3D model of the buildings at LOD1. The generated model was saved in the desired output format, CityGML.
	
	\subsection{\textcolor{red}{Feature Extraction for Air Temperature Downscaling}}\label{sec_4.4}
	After creating a 3D model, morphological and meteorological parameters that can be used to downscale air temperature are extracted. Air temperature is a complex phenomenon influenced by various factors. Previous studies have highlighted the significant role of urban morphology in shaping local air temperature patterns \citep{RN10, RN108, RN192}. In this study, urban morphology parameters were employed to better understand the relationship between urban form and air temperature. These parameters included SVF, H, BR, WALL, P, CCR, DW, and DP. The meteorological parameters include daily SR, daily WS, and daily RH. The variables SR, RH, and Wind are dynamic, meaning they change over time, while the other variables are static. All mentioned features are extracted at a resolution of 5 m. The basis for selecting this resolution involves a trade-off between the level of noise and the level of detail in the morphological features extracted from LiDAR data. At very high resolutions, such as 1 m, the level of noise in the LiDAR data is significantly high, which affects the extraction of high-quality features. At medium resolutions, such as 10 m, the structure of the buildings becomes simplified, which also reduces the quality of the morphological features. Therefore, a resolution of 5 m appears to be an appropriate intermediate value for extracting suitable morphological features. Additionally, in previous studies, a resolution of 5 m has been used in the Amsterdam region for estimating meteorological parameters \citep{Schmitz2019-pj}.
	
\begin{figure}[t]
		\centering
		{\includegraphics[width=\linewidth]{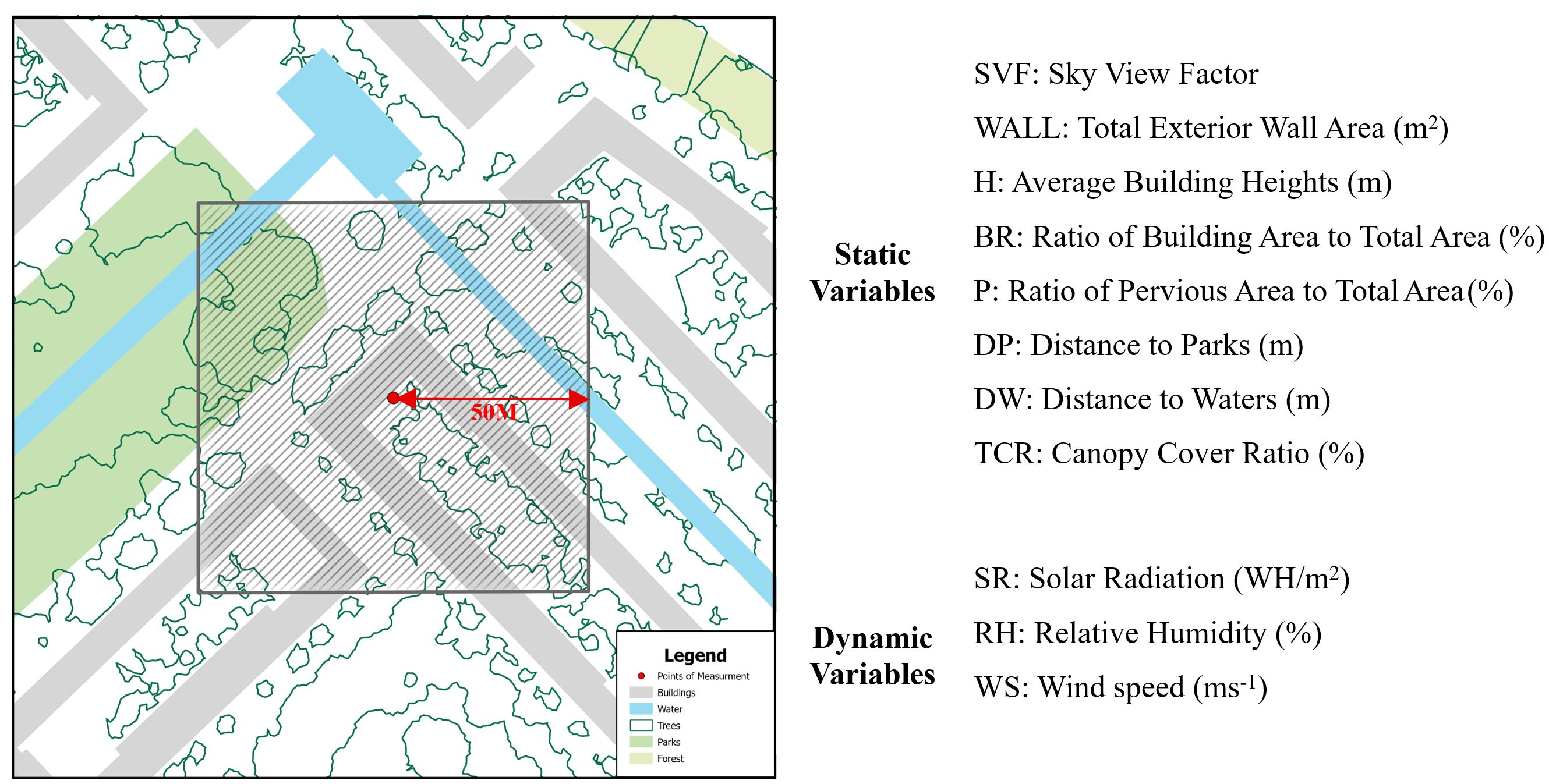}}
		\vspace{0.05cm}
		\caption{The influence area with a radius of 50 m was considered for extracting features such as those listed above.}
		\label{buffer_fig}
	\end{figure}

	According to Fig.\ref{buffer_fig}, to calculate the features for the point of interest, an influence area with a specified radius is considered, which refers to the area where the surrounding environment affects the air temperature. In this study, three radii of 25 m, 50 m, and 75 m were considered. As will be discussed in Section \ref{sec_5.2}, considering an influence area with a radius of 50 m will yield the best results, which is consistent with previous studies in this field \citep{RN108, RN107}. Each feature is briefly explained in the following, and further details on their calculation methods are provided in Appendix \ref{appendix_features}.
	
	\textcolor{red}{Sky View Factor (SVF) measures the visible sky from a ground point, ranging from 0 (obstructed) to 1 (unobstructed) \citep{RN154, RN158, RN157, RN156}. It is used to study the effects of urban geometry on air temperature \citep{RN156, RN154, RN163, RN160, RN161, RN159}. The UMEP plugin was used to calculate SVF \citep{RN75}.
	Building height (H) affects the urban microclimate by influencing airflow and air temperature, with taller buildings increasing temperature and shorter ones reducing it. 
	Building Coverage Ratio (BR) indicates the proportion of building area to the total area, where higher BR can lead to higher ambient air temperatures \citep{RN197}. 
	The total exterior wall area (Wall) indicates building density \citep{RN108}, which was calculated with the help of the CityGML model (created in Section \ref{sec_4.3}). 
	Parks and water bodies affect the ambient air temperature of the surrounding area \citep{RN196, RN195}. Distances to parks (DP) and water bodies (DW) were calculated to measure their cooling effects on surrounding areas.
	Increasing tree cover can lead to air temperature reduction \citep{RN195}. Therefore, the Canopy Cover Ratio (CCR) index can  effectively downscale air temperature.  
	Pervious surfaces, such as water bodies and green areas, moderate urban air temperature. The ratios of these surfaces were specified using OSM data (water, parks, and forest layers) and the obtained tree coverage.
	Solar radiation (SR) absorbed by urban surfaces converts to heat, raising air temperature. Factors like building density, orientation, and green spaces affect SR levels. SR was calculated as the daily average over four months \citep{RN229}.
	Wind speed affects air temperature variably, either cooling or warming it. Wind data from the UrbClim model were averaged, showing minimum and maximum daily values, and resampled from 100 m to 5 m resolution.
	Relative humidity affects the perception of air temperature, making it feel warmer at higher levels and cooler at lower levels \citep{RN230}. RH data were obtained from the UrbClim model, which is similar to wind speed data.}

	\subsection{Development of a Machine Learning Model for Air Temperature Downscaling}
	\subsubsection{\textcolor{red}{Regression Models}}\label{sec_4.5.1}
	Air temperature is a complex variable influenced by multiple factors, and the impact of each factor may not be straightforward. Because of the complex relationship between air temperature and the mentioned features, a more sophisticated approach, such as machine learning, can be suitable for modeling and estimating air temperature. These algorithms can learn complex relationships between variables and understand nonlinear relationships. They also can combine multiple variables simultaneously for more accurate temperature estimation.

	This section describes a machine learning model developed to downscale the air temperature from the UrbClim model from 100 m to 5 m resolution using urban morphological features and meteorological parameters (introduced in Section \ref{sec_4.4}), coordinates (X, Y), and the day of the year (DOY) as predictor variables. These variables can help the model capture air temperature variations at different temporal and spatial scales. The goal is to establish a correlation between these predictors and air temperature. To achieve this, it is necessary to determine an appropriate model that can effectively demonstrate the relationship between these variables. Machine learning techniques provide a powerful approach for estimating this correlation model. These models can effectively discover the relationships between input variables and the air temperature parameter and provide accurate estimations. For this purpose, several machine learning regression models are examined, including Support Vector Regression (SVR), Random Forest (RF), Extremely Randomized Trees or Extra Trees (ET), XGBoost, and LightGBM. \textcolor{red}{Table \ref{table_MLmodels} summarizes the main characteristics of the mentioned models.}
	
	\begin{table}[t]
		\centering
		\caption{\textcolor{red}{Description of machine learning algorithms used in this study}}
		\label{table_MLmodels}
		\begin{adjustbox}{width=\textwidth}
			\begin{tabular}{c|c|c}
			\toprule
			\textbf{Algorithm} & \textbf{Description} & \textbf{Key Features} \\
			\midrule
			SVR & Generalized version of SVM for regression problems \citep{RN164, RN165}. & \makecell{Aims to find a function to estimate the relationship between input variables \\ and a continuous target variable by minimizing the prediction error.} \\[.5cm]
			RF & \makecell{Ensemble learning method combining multiple decision \\ trees to solve regression and classification problems \citep{RN120}.} & \makecell{Uses bagging to train multiple decision trees independently \\ and averages their outputs for the final prediction.} \\[.5cm]
			ET & \makecell{Ensemble learning method similar to RF, \\ combining multiple decision trees \citep{RN121}.} & \makecell{Uses bagging but randomly selects thresholds \\ for node splitting, improving speed.} \\[.5cm]
			XGBoost & \makecell{Extreme Gradient Boosting, known for high speed and \\ accuracy in complex regression and classification tasks \citep{RN115}.} & \makecell{Uses gradient boosting decision trees, updating model weights with gradients,\\ leading to a stronger and less biased model over time \citep{RN172, RN171}.} \\[.5cm]
			LightGBM & \makecell{Light Gradient Boosting Machine, a gradient-boosting \\ algorithm based on decision trees \citep{RN117}.} & \makecell{Grows trees leaf-wise rather than level-wise, offering \\ high training speed, accuracy, and lower memory usage.} \\
			\bottomrule
			\end{tabular}
		\end{adjustbox}
	\end{table}
	
	\subsubsection{\textcolor{red}{Regression Model Deployment and Performance Evaluation}} \label{sec_4.5.2}
	
	The implementation and accuracy assessment of the models were conducted using various metrics including Root Mean Squared Error (RMSE), Mean Absolute Error (MAE), and coefficient of determination (R²) \citep{RN191} in two stages. In the first stage, various regression models were trained and evaluated using a smaller amount of training data (65 spatial locations for training and 5 locations for testing) to select the top-performing model. For each training location, for four months (January, May, July, and October totaling 124 days), 11 urban morphological features were extracted from LiDAR data considering a 100 m buffer. These features, along with the X and Y of locations and DOY, were the input parameters for the machine learning models. Correspondingly, the air temperature at each location was extracted from UrbClim data and used as the target variable. In total, 8680 training samples (70 locations × 124 days) of 14 input features and one target variable (air temperature) were obtained. These training samples were then used to train various regression models to discover the relationship between the air temperature parameter and the input features.
	
	Before training the regression models, preprocessing was necessary. Since each input variable had different ranges, they were normalized to standardize the range of variation. To optimize the performance of machine learning algorithms, the hyperparameters of each algorithm were tuned using Grid Search and Bayesian optimization methods, with the K-Fold cross-validation technique (K = 5). Grid Search provided better results for most algorithms but required longer execution times, whereas Bayesian optimization yielded better results for the LightGBM model. The tuned hyperparameters for each model are presented in Appendix \ref{appendix_hp_table}, Table \ref{table_hp}.
	
	After training the model and determining the optimal hyperparameters, the accuracy of air temperature estimation by the developed regression models was evaluated relative to the reference air temperature (obtained from the UrbClim model) at test points. Based on the results of this stage, the best model (LightGBM) was selected. To improve generalization, the training process for the LightGBM model was repeated with a larger number of training data (1143 locations). The extent of training locations is shown in Fig.\ref{study_area}.
	
	The final tuned LightGBM model was then used to downscale air temperature, utilizing the 5 m resolution features (input parameters) to estimate the air temperature at 5 m resolution. The accuracy of the downscaled air temperature was evaluated both absolutely (compared to the ground station) and relatively (compared to the UrbClim model). In the absolute evaluation, the mean, maximum, and minimum of the 5 m resolution estimated air temperature were compared with observed values at the Schiphol ground meteorological station for three consecutive days in each of the four months, totaling 12 days. Additionally, for a more detailed examination, the air temperature estimated by the UrbClim model at the ground station location was compared with the observed values at that ground station.
	
	Due to the limited ground reference data, in addition to the absolute evaluation, a relative accuracy evaluation of the 5 m air temperature estimation was also performed. For this purpose, 18009 locations were considered as relative evaluation points. In these locations (similar to the ground station), the air temperature from the UrbClim model and the corresponding output from the downscaling process were extracted. It was assumed that the air temperature estimation by the UrbClim model is highly accurate, and its output can be used as a reference for evaluating the accuracy of the downscaled air temperature estimation. More precisely, the output of the UrbClim model was considered similar to ground measurements (pseudo-ground measurement) located 100 m apart. This relative evaluation provides a more accurate assessment of the downscaled estimation results due to the high number of measurements at different locations.
	
	\section{Results}\label{sec_5}
	\subsection{Results of Building Footprints Detection and 3D Modelling}
\begin{figure}[t]
		\centering
		{\includegraphics[width=0.6\linewidth]{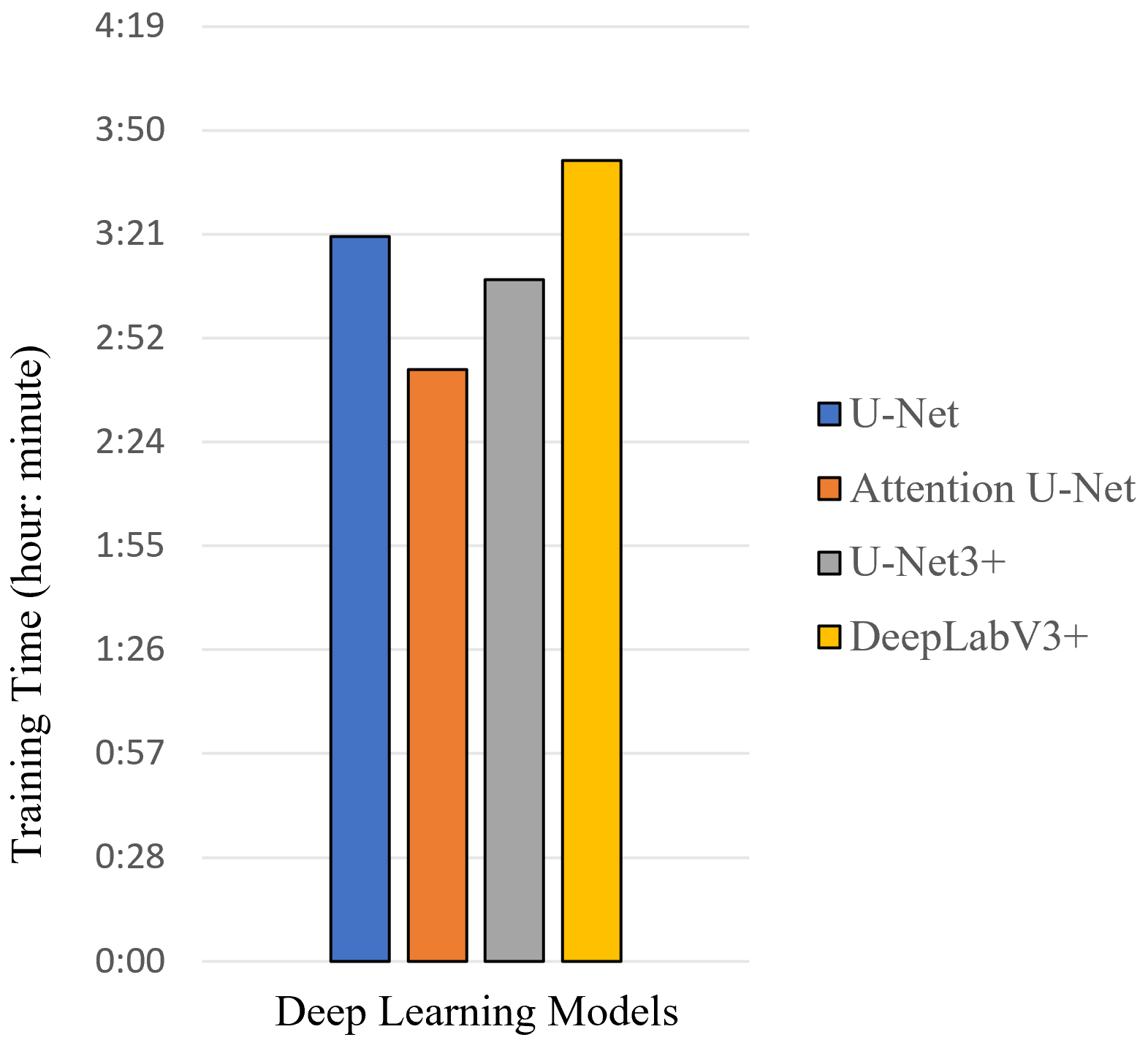}}
		\vspace{0.05cm}
		\caption{Comparing training time of deep semantic models for building footprint extraction}
		\label{train_time_chart}
	\end{figure}

	The footprints of buildings were extracted from deep learning models; U-Net, Attention U-Net, U-Net3+, and DeepLab V3+ according to the strategy described in Section \ref{sec_4.2}. The performance of each model on the Miami-Dade test dataset are presented in Table \ref{table_3}. According to Table \ref{table_3}, all models exhibited almost the same performance across various metrics. The U-Net3+ model followed by the Attention U-Net model with almost similar results showed the best performance. The U-Net and DeepLabV3+ models also achieved acceptable results, although they had slightly weaker performance and slower training speed compared to the other two models (see Fig.\ref{train_time_chart}).
	
	\begin{table}[t]
		\centering
		\caption{Performance of deep semantic models on Miami-Dade test dataset.}
		\begin{tabularx}{\textwidth}{c|X|X|X|X|X}
			Model & IoU & F1 Score & Accuracy & Precision & Recall \\
			\toprule
			U-Net & 0.841 & 0.913 & 0.967 & 0.899 & 0.928 \\
			Attention U-Net & 0.852 & 0.920 & 0.970 & \textbf{0.908} & 0.933 \\
			U-Net3+ & \textbf{0.853} & \textbf{0.921} & \textbf{0.970} & 0.904 & \textbf{0.938} \\
			DeepLab V3+ & 0.842 & 0.914 & 0.968 & 0.907 & 0.922 \\
		\end{tabularx}
		\label{table_3}
	\end{table}

	The top-performing models, U-Net3+ and Attention U-Net, were selected for training on the Amsterdam dataset, and were fine-tuned using transfer learning for more accurate extraction of footprints in the desired region. The performance of the models on the Amsterdam test dataset are presented in Table \ref{table_4}. Overall, both models provided acceptable results for the Amsterdam dataset. However, the U-Net3+ model outperformed the Attention U-Net model with an IoU of 0.83, F1 score of 0.91, accuracy of 0.96, precision of 0.95, and recall of 0.88. Although the difference in precision between the two models is very small (U-Net3+ with a precision of 0.9453 and Attention U-Net with a precision of 0.9431), indicating that both models are capable of correctly identifying building footprints (True Positive) with a high level of accuracy, the U-Net3+ model performed better according to other metrics. Considering the negligible difference in training time between the two models and the higher accuracy of the U-Net3+ model (before and after transfer learning), this model was selected as the final model for extracting building footprints from LiDAR data at the city level in Amsterdam. 
	
	\begin{table}[t]
		\centering
		\caption{Performance evaluation of deep semantic models for Amsterdam Data (after applying transfer learning)}
		\begin{tabularx}{\textwidth}{c|X|X|X|X|X}
			\centering Model & IoU & F1 Score & Accuracy & Precision & Recall \\
			\toprule
			U-Net3+ & \textbf{0.833} & \textbf{0.909} & \textbf{0.961} & \textbf{0.945} & \textbf{0.875} \\
			Attention U-Net & 0.814 & 0.898 & 0.956 & 0.943 & 0.856 \\
		\end{tabularx}
		\label{table_4}
	\end{table}
	
\begin{figure}[t]
		\centering
		{\includegraphics[width=\linewidth]{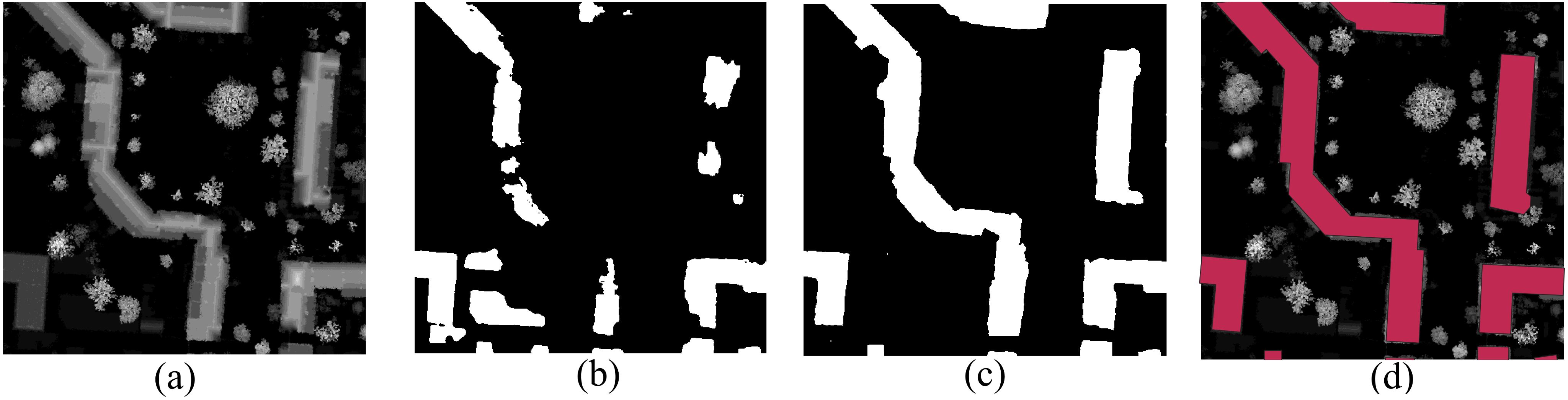}}
		\vspace{0.05cm}
		\caption{An example sample of detected building footprints by U-Net3+ model before and after applying transfer learning: (a) LiDAR DSM, (b) Binary image representing building before applying transfer learning, (c) Binary image representing building after applying transfer learning, (d) Footprints after post-processing overlaid on DSM}
		\label{footprint_sample_fig}
	\end{figure}

	Fig.\ref{footprint_sample_fig} shows an example LiDAR sample along with building footprints extracted from it by U-Net3+ before and after transfer learning. Fig.\ref{footprint_sample_fig}.b depicts the initial binary raster output of the model. The U-Net3+ model was able to detect buildings in high-density areas, buildings with complex shapes, and buildings with small areas. After applying the post-processing methods described in Section \ref{sec_4.2.5}, the final building footprints were refined, as shown in Fig.\ref{footprint_sample_fig}.d. As shown, the post-processing techniques reduced noise and eliminated undesirable distortions in the footprints. Comparing Fig.\ref{footprint_sample_fig}.b and Fig.\ref{footprint_sample_fig}.c demonstrates the importance of employing transfer learning to improve the performance of the U-Net3+ model. These figures depict the results of building detection by the U-Net3+ model before and after fine-tuning. As a result, transfer learning and fine-tuning of the deep learning model improve the accuracy of building footprint extraction in the study area.
	
\begin{figure}[!ht]
		\centering
		\subfloat{{\includegraphics[width=0.75\linewidth]{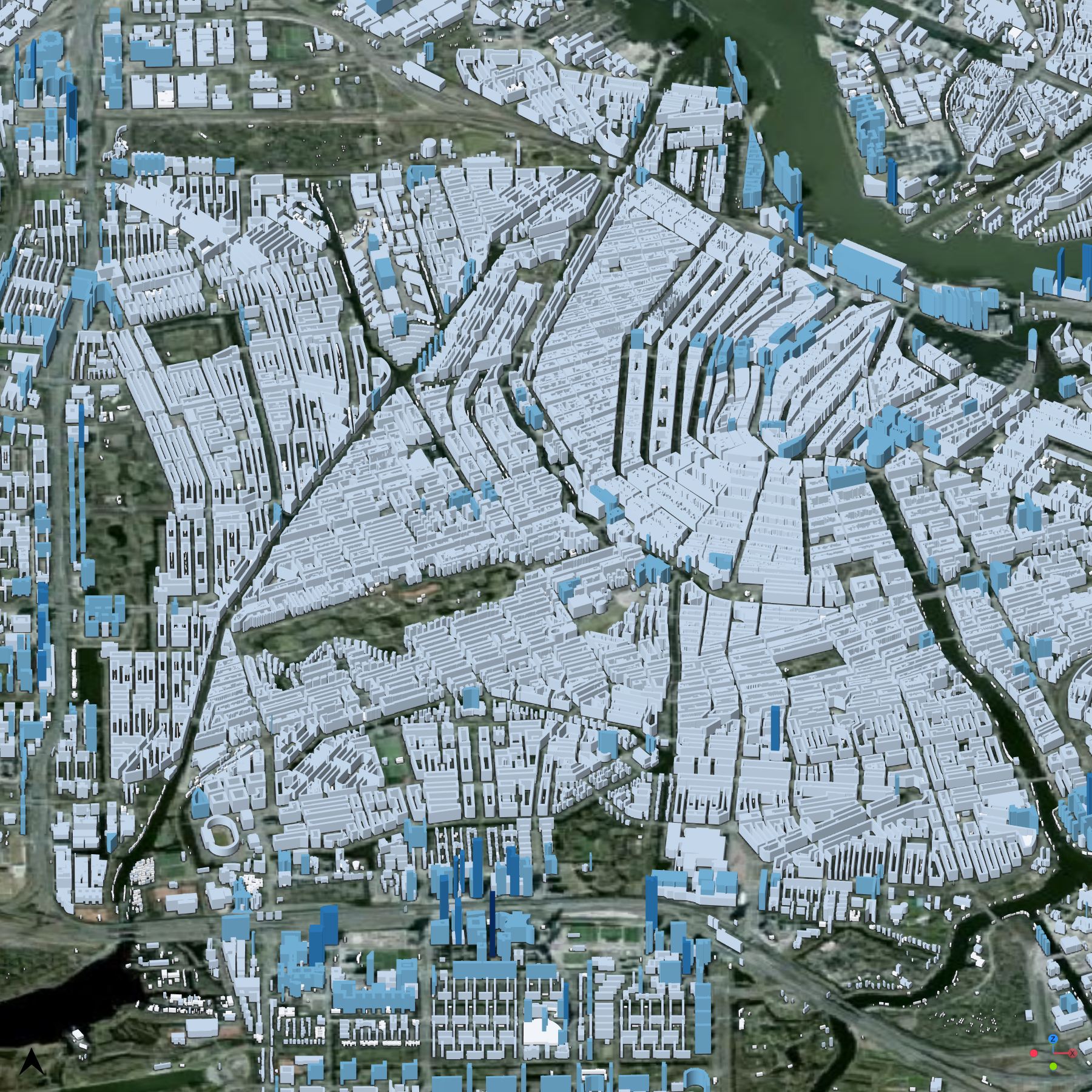}}}%
		\qquad
		\subfloat{{\includegraphics[width=0.15\linewidth]{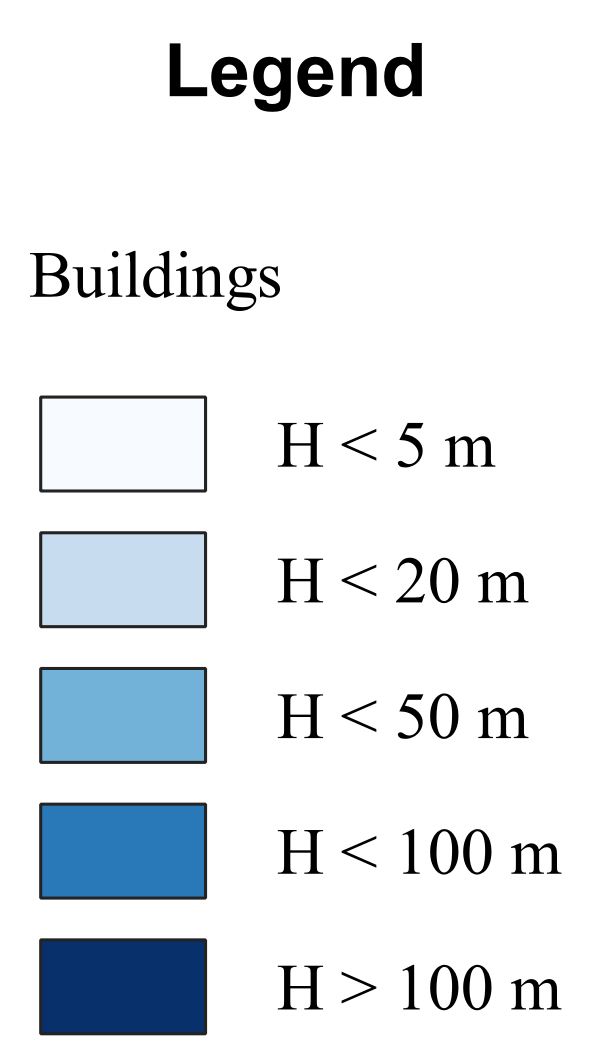} }}%
		\caption{Height-based colorization of a 3D building model (LOD1) of downtown of Amsterdam}
		\label{3dmodel_H_fig}
	\end{figure}

	According to the explanations given in Section \ref{sec_4.3}, using the final building footprints, a LOD 1 as a 3D model of the buildings was created for the entire city of Amsterdam \citep{RN130}. Fig.\ref{3dmodel_H_fig} visualizes the created 3D model for a part of the city center with a color-coded representation based on building heights. As observed, the city center consists of dense and low-rise buildings, while in the suburbs, the heights of the buildings increase and the building density decreases.
	
	\subsection{Result of Feature Extraction}\label{sec_5.2}
\begin{figure}[t]
		\centering
		{\includegraphics[width=0.95\linewidth]{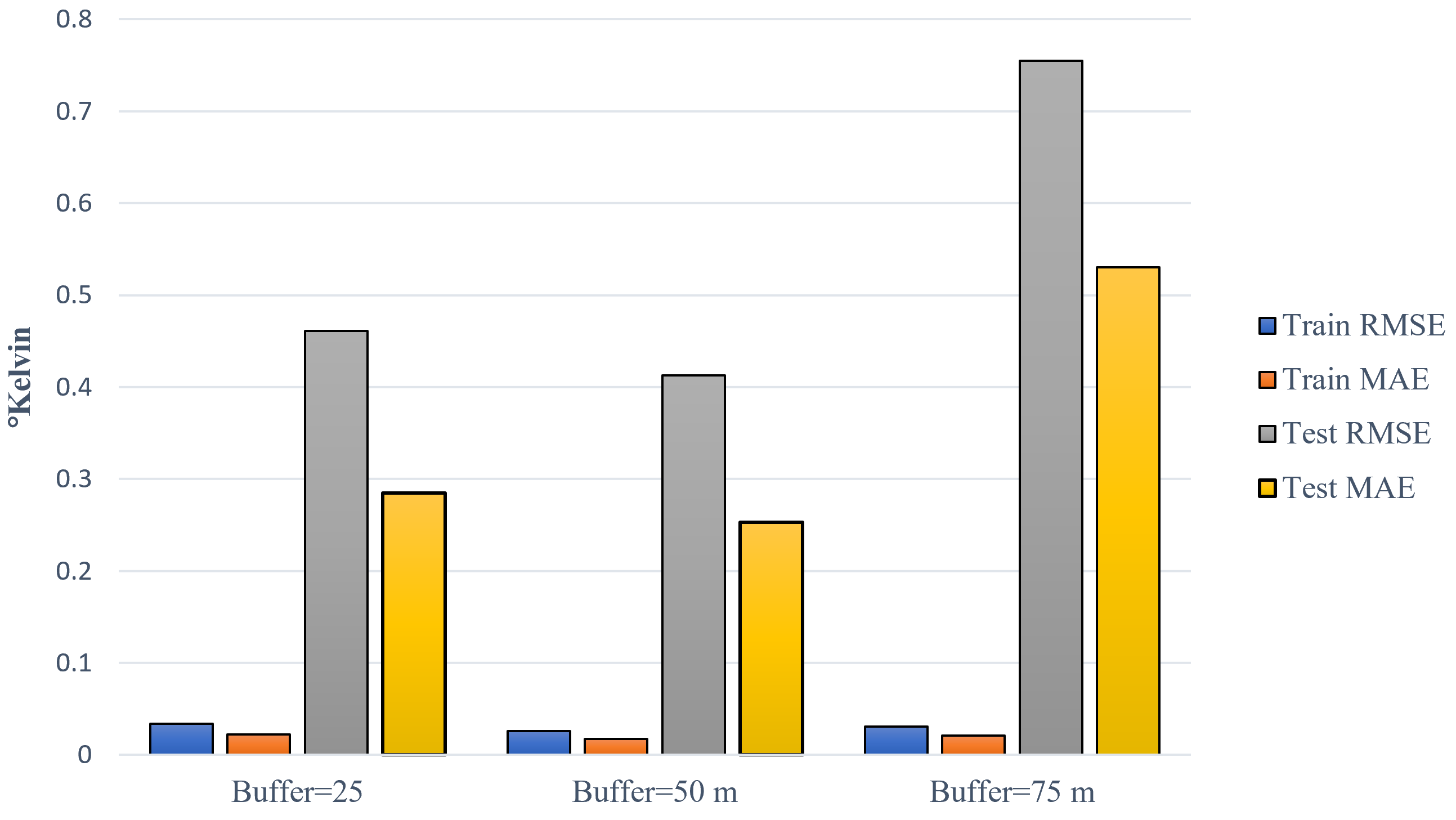}}
		\vspace{0.05cm}
		\caption{Impact of buffer sizes on air temperature downscaling using LightGBM}
		\label{buffer_chart}
	\end{figure}

	According to the description in Section \ref{sec_4.4}, features were extracted within an influence radius (buffer). In this study, a comparison was made between three different buffers with sizes of 25 m, 50 m, and 75 m as the influence area for calculating each feature, and the impact of buffer size on air temperature estimation accuracy (using the LightGBM model) was investigated. The results of this analysis are shown in Fig.\ref{buffer_chart}, indicating that considering a 50 m buffer yields the best results. The reference air temperature data extracted from the UrbClim model are at a resolution of 100 m, meaning that each location in this model represents an average air temperature within a 50 m buffer. This can be a reason for the superiority of using a 50 m buffer in air temperature downscaling. A 25 m buffer can increase the possibility of noise occurrence by providing more details, thus reducing the accuracy of air temperature estimation. A 75 m buffer extracts fewer details and may result in the loss of some morphological features, leading to a decrease in air temperature estimation accuracy. Therefore, a 50 m buffer creates a suitable balance between details and noise level, and it can provide more accurate results.
	
\begin{figure}[!t]
		\centering
		{\includegraphics[width=0.95\linewidth]{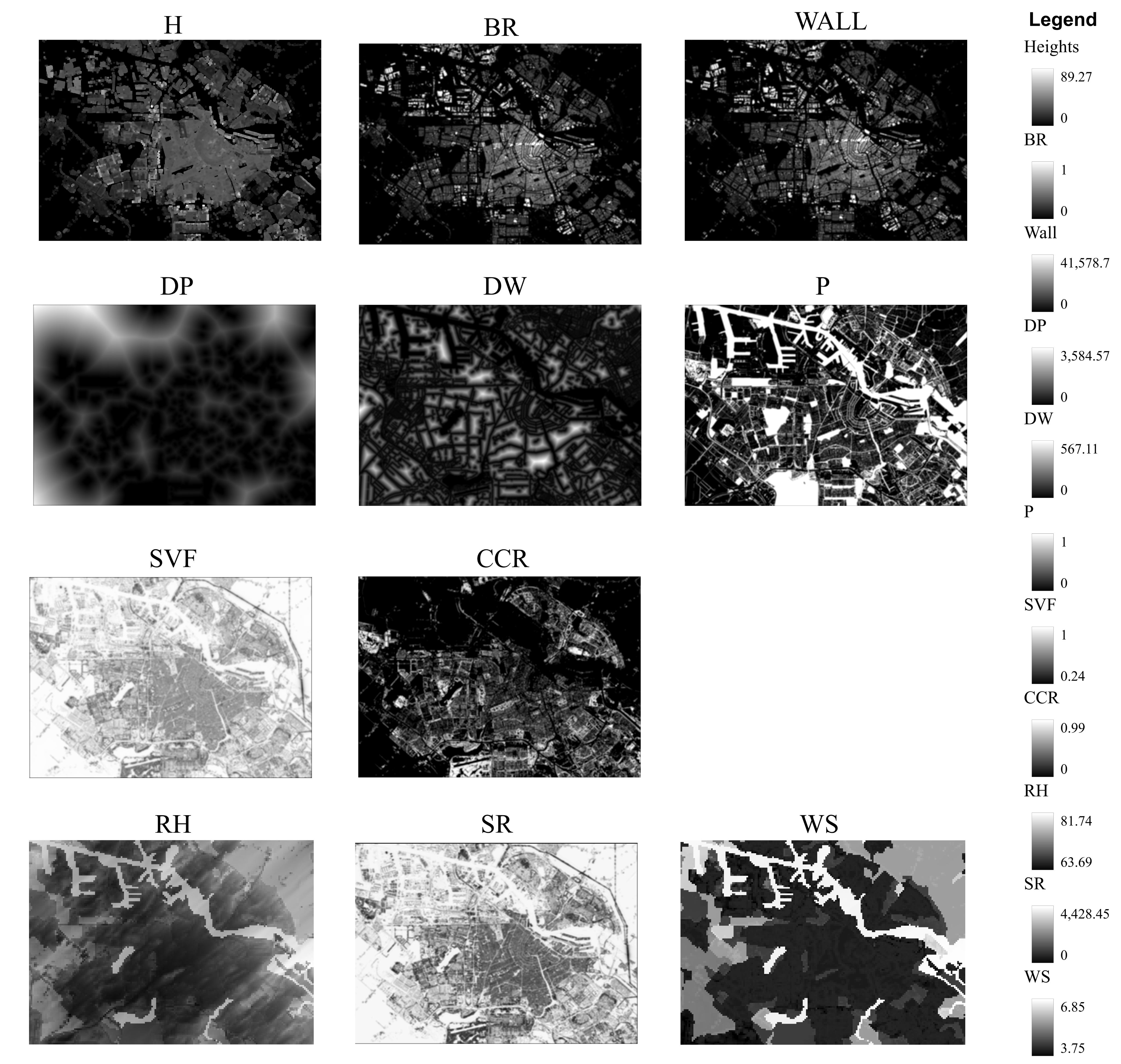}}
		\vspace{0.05cm}
		\caption{Visualization of extracted features used for air temperature downscaling over the study area (5 m resolution). Dynamic features including RH, SR, and WS are rasterized for $4^{th}$ of May, 2017.}
		\label{features_fig}
	\end{figure}

	As mentioned, except for the three dynamic features of WS, RH, and SR, all other features are static. In Fig.\ref{features_fig}, the extracted features are shown. Dynamic features are shown for May 4, 2017, as a sample. The impact of building morphology on the 3D model-extracted features is clearly reflected. For example, in the city center where building density is higher, the H, BR, and WALL values are higher. As we move away from the city center, the building density decreases, and the relevant values decrease as well. Since Amsterdam is situated on a network of canals, the DW does not have high values. The CCR value is higher in the outskirts of the city due to the presence of more green spaces and parks, while this value decreases in the central core of the city due to high building density. The SVF is higher in open spaces such as parks, where obstacles like tall buildings are infrequent. Additionally, SVF values are also higher in low-rise areas compared to high-rise ones. In contrast, forested areas with high tree density have limited sky view, resulting in lower SVF values. Narrow streets and the presence of buildings on both sides also contribute to a decrease in SVF value. In Amsterdam's city center, where streets are narrow and buildings are closely spaced, SVF is lower. On rooftops, SVF is higher due to fewer obstacles. However, due to the averaging and resolution reduction in SVF calculation, the influence of streets between buildings is more pronounced, leading to a decrease in SVF value. In Amsterdam, due to its proximity to the North Sea, the RH is generally high. In general, areas closer to major rivers tend to have higher RH, while those farther away from these rivers, especially in downtown, where the distance from major rivers increases and building coverage is higher, tend to have lower RH. The amount of SR in densely urban areas (city center) may be lower due to the shadow cast by tall buildings. Similarly, areas with low SVF have lower SR due to the presence of obstacles. Areas with more open spaces (higher SVF) have higher SR due to fewer obstacles to prevent sunlight. High WS can help dissipate clouds, which act as barriers to solar radiation, so the amount of SR may increase in areas with higher WS.

	\subsection{Result of Regression by Machine Learning Techniques }
	In the first phase, the regression models mentioned in Section \ref{sec_4.5.1} were implemented to establish a relationship between the extracted features and air temperature in a small area in the city center. Table \ref{table_5} represents the performance of five regression models, using metrics such as RMSE, MAE, and $R^2$ for training and test data. The results indicated that the LightGBM model estimated the air temperature with the highest accuracy compared to other models, with an RMSE of 0.15\textdegree K, MAE of 0.06\textdegree K, and $R^2$ of 0.99 for the test data. On the other hand, the SVR model exhibited the weakest performance, with an RMSE of 2.07\textdegree K, MAE of 1.5\textdegree K, and lower $R^2$ of 0.89. The high $R^2$ metrics indicated that the predictor variables effectively explain air temperature variations. In summary, LightGBM is ideal for large-scale air temperature downscaling due to its superior performance, faster training, and efficient, accurate leaf-wise tree growth.
	
	\begin{table*}[!t]
		\centering \footnotesize
		\caption{Comparison of machine learning models for estimating mean air temperature in a small region of interest}
		\label{table_5}
		\begin{tabular}{c |cc|ccc}
			
			&  \multicolumn{2}{c|}{Train}& \multicolumn{3}{c}{Test} \\
			Method & RMSE (\textdegree K)  & MAE (\textdegree K)  & RMSE (\textdegree K)  & MAE (\textdegree K)  & $ R^{2} $ \\
			\toprule
			XGBoost &	0.135&	0.066&	0.454&	0.297&	0.99 \\
			RF &	0.254&	0.129&	0.244&	0.129&	0.99 \\
			SVR &	2.058&	1.491&	2.07&	1.501&	0.89 \\
			LightGBM &	\textbf{0.025}&	\textbf{0.015}&	\textbf{0.148}&	\textbf{0.062}&	\textbf{0.99}\\
			ET &	0.392&	0.265&	0.387&	0.262&	0.99 \\
		\end{tabular} 	
	\end{table*}

	\subsection{\textcolor{red}{Result of Air Temperature Downscaling}}
	The initial evaluation of the results showed that the LightGBM model with its hyperparameters tuned as detailed in Table \ref{table_hp}, outperformed other models in estimating air temperature. Consequently, this model was used as the final tuned model in the air temperature downscaling process. 
	
	
	Table \ref{table_stat} provides the absolute assessment, compared the estimated air temperatures at a 5 m resolution with the observed air temperatures at the Schiphol station. The evaluation used metrics such as MAE, RMSE, STD, and Median for 12 days across different seasons. To better assess the results, the relative difference in the errors of the downscaling and UrbClim model outputs compared to the ground observations was also calculated for each metric and represented as a percentage ($ \text{diff} =  \frac{\left|\text{Downscaled-UrbClim}\right| }{\text{UrbClim}} $). The results indicate that the difference between the UrbClim model air temperature and the ground station is slightly different from that between the LightGBM downscaled air temperature and the ground station. Considering the median metric, the difference is at worst 15.8\% for minimum air temperature and at best 7.5\%  for maximum air temperature, indicating the success of the developed downscaling process. However, it can be concluded that the downscaling process for the minimum air temperature performed weaker, likely due to the weaker performance of the UrbClim model in estimating minimum air temperatures as well. Overall, the results show that the downscaling process using LightGBM has increased the resolution to 5 m while striving to maintain the accuracy of the initial air temperature estimation provided by the UrbClim model.

	\begin{table}[t]
		\centering
		\caption{Absolute Performance Evaluation of air Temperature Estimation: Downscaled and UrbClim vs. Ground Truth}
		\label{table_stat}
		\begin{adjustbox}{width=\textwidth}
			\begin{tabular}{cccc|ccc|ccc}
				\toprule
				& \multicolumn{3}{c}{Average Air Temperature (°K)} & \multicolumn{3}{c}{Minimum Air Temperature (°K)} & \multicolumn{3}{c}{Maximum Air Temperature} \\
				\cmidrule(lr){2-4} \cmidrule(lr){5-7} \cmidrule(lr){8-10}
				& \makecell{UrbClim \\ (100m)} & \makecell{Downscaled \\ (5m)} & diff & \makecell{UrbClim \\ (100m)} & \makecell{Downscaled \\ (5m)} & diff & \makecell{UrbClim \\ (100m)} & \makecell{Downscaled \\ (5m)} & diff \\
				\midrule
				MAE & 1.02 & 1.213 & 15.9\% & 2.2 & 2.564 & 14.2\% & 0.763 & 0.739 & 3.0\% \\
				Median & 0.915 & 0.907 & 9.0\% & 1.841 & 2.186 & 15.8\% & -0.201 & -0.187 & 7.5\% \\
				STD & 0.633 & 0.834 & 24.1\% & 1.325 & 1.48 & 11.0\% & 0.966 & 1.034 & 6.6\% \\
				RMSE & 1.2 & 1.472 & 18.5\% & 2.569 & 2.96 & 13.2\% & 0.977 & 1.044 & 6.4\% \\
				\bottomrule
			\end{tabular}
		\end{adjustbox}
	\end{table}
	
\begin{figure}[t]
		\centering
		{\includegraphics[width=0.6\linewidth]{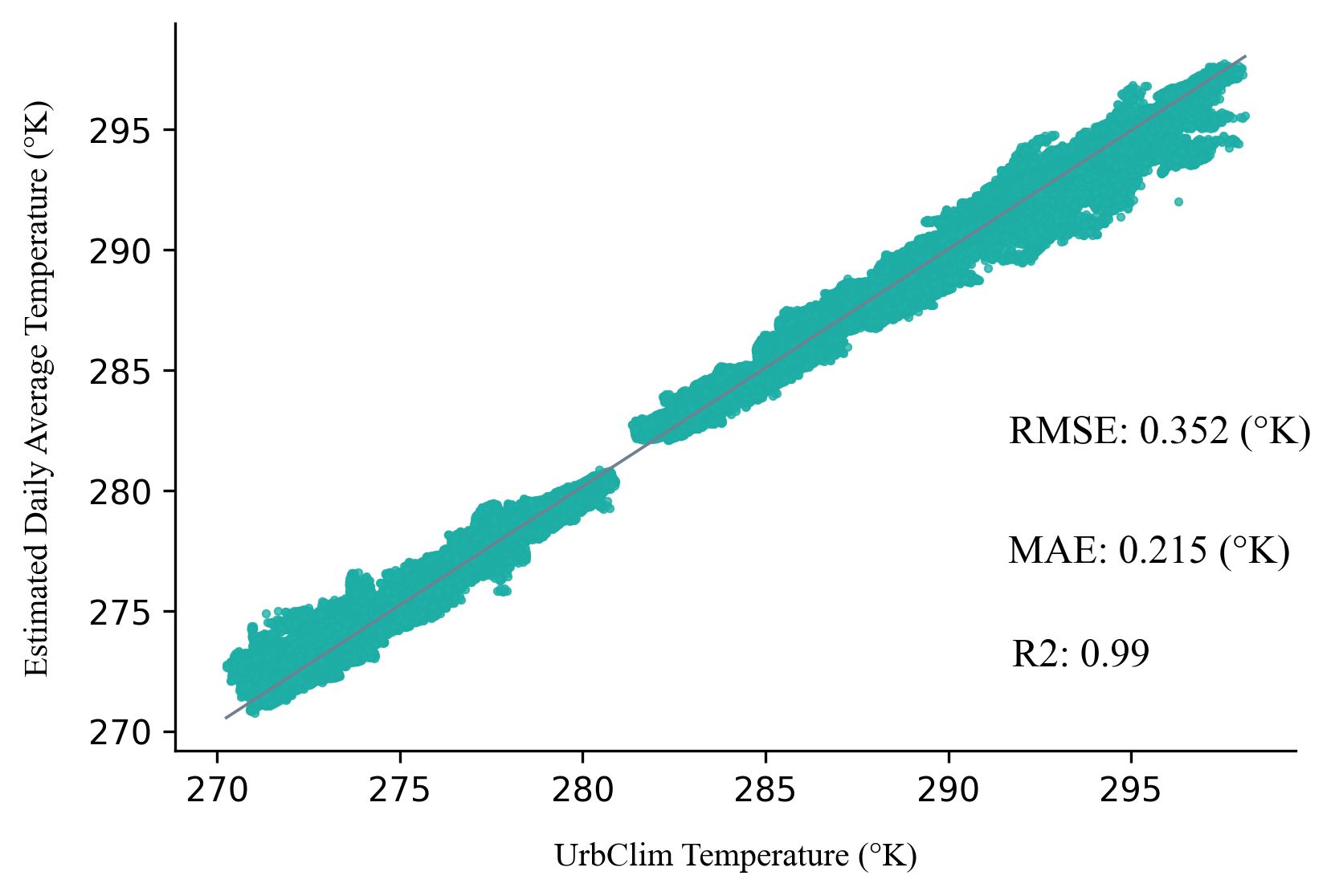}}
		\vspace{0.05cm}
		\caption{Performance of LightGBM model for estimating daily average air temperature in Amsterdam}
		\label{regplot_fig}
	\end{figure}

	In addition to absolute evaluation, a relative evaluation of the air temperature downscaling accuracy was performed using the UrbClim model itself. For this purpose, some points where air temperature was estimated by the model were excluded from the entire downscaling process (18,009 points located in the study area), and were used in the final evaluation of the downscaling. Fig.\ref{regplot_fig} illustrates the scatter distribution and correlation between the estimated daily average air temperature obtained through downscaling and the reference air temperature derived from the UrbClim model. The plot shows a high correlation ($R^2$ = 0.997) between the estimated air temperature and the reference air temperature with the RMSE and MAE of 0.352\textdegree K and 0.215\textdegree K, respectively, for the test points, that demonstrated the downscaling process was achieved in the Amsterdam region with highly accurate estimations. Figures S2 and S3 present the scatter distributions for the daily maximum and daily minimum air temperatures, with RMSE and MAE values for minimum air temperature reported as 0.477 and 0.303, and for maximum air temperature as 0.426 and 0.238.
	
\begin{figure}[!tb]
		\centering
		{\includegraphics[width=1\linewidth]{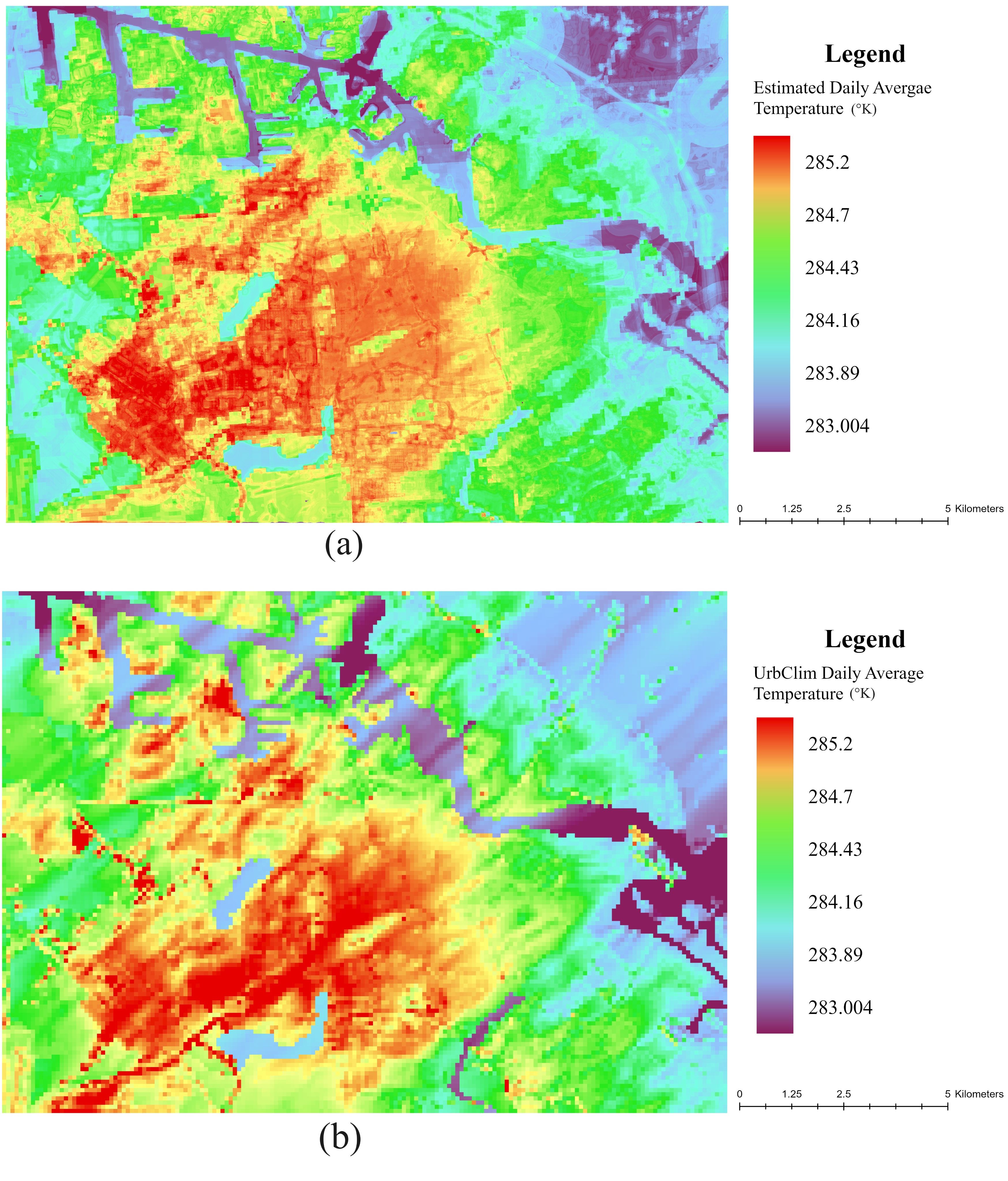}}
		\vspace{0.05cm}
		\caption{Mean air temperature of Amsterdam for 4th of May, 2017: (a) estimated air temperature in 5 m resolution (b) UrbClim air temperature in 100 m resolution}
		\label{TempMap_fig}
	\end{figure}

	Furthermore, Fig.\ref{TempMap_fig}.a presents a high-resolution (5 m) map of the average daily air temperature for May 4th, while Fig.\ref{TempMap_fig}.b shows the average air temperature modeled by UrbClim at a 100 m resolution for the same day. Comparing these maps reveals that the 5 m resolution downscaled air temperature provides more detailed than the 100 m resolution UrbClim model. In other words, compared to the coarse air temperature representation at a 100 m resolution, the 5 m resolution captures air temperature variations at a larger scale and depicts local air temperature patterns more accurately. For instance, the air temperature patterns in the vicinity of the canals, as well as the effect of streets and buildings, particularly in the city center, are clearly visible in Fig.\ref{TempMap_fig}.a. 
	Despite the resolution difference, the overall air temperature patterns are similar, indicating the machine learning model has successfully preserved the overall air temperature trend at a 5 m resolution during the downscaling process. The range of air temperature variations in the high resolution map (282.98\textdegree to 285.25\textdegree K) is nearly equivalent to the range of air temperature variations in the 100 m resolution map (283.25\textdegree to 285.48\textdegree K).
	
\begin{figure}[!tb]
		\centering
		{\includegraphics[width=0.85\linewidth]{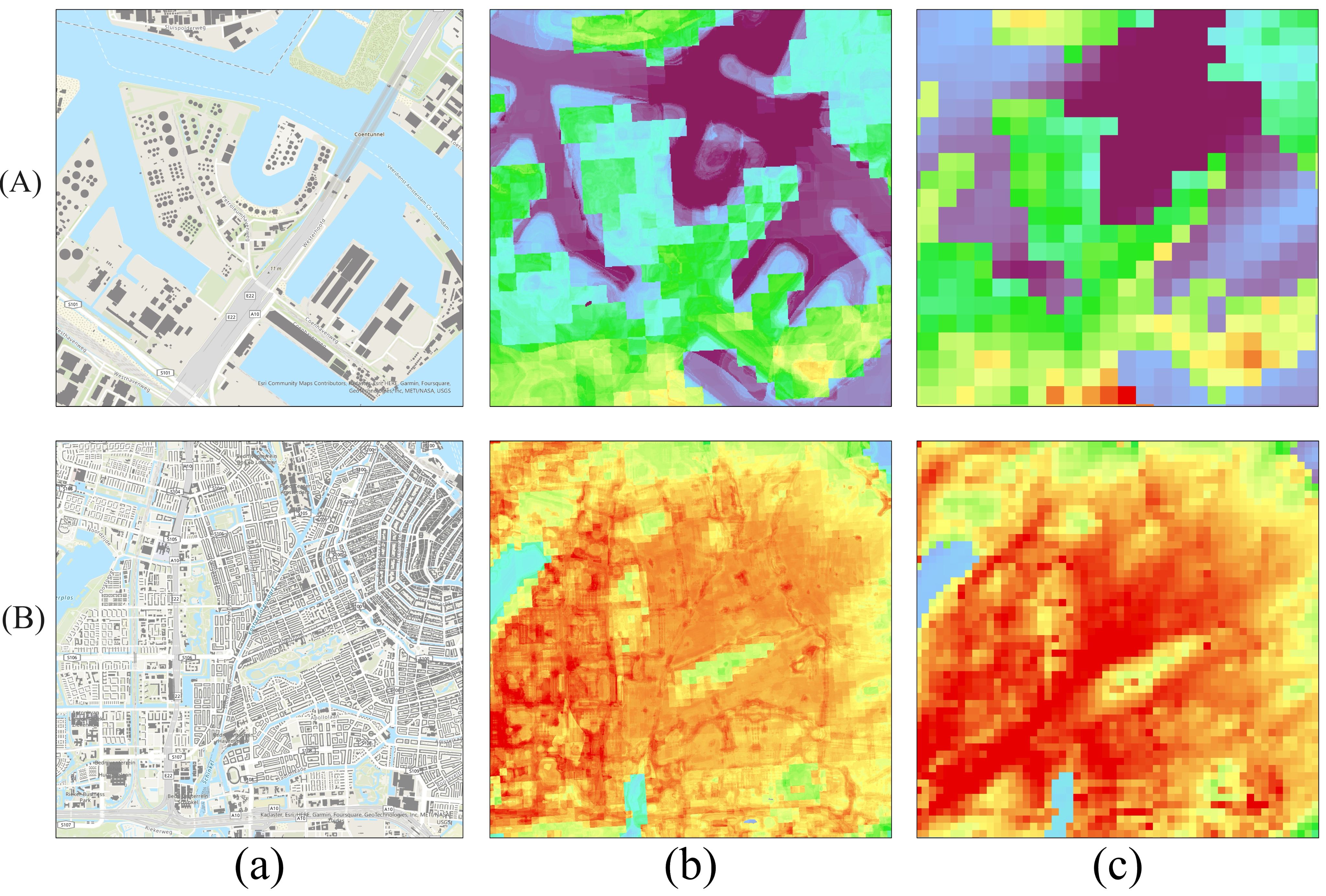}}
		\vspace{0.05cm}
		\caption{Comparing high resolution (5 m) and low resolution air temperature maps (100 m) to reveal localized patterns in Amsterdam: (a) topographic map with building footprint overlay (b) estimated air temperature map (5 m) (c) UrbClim air temperature map (100 m)}
		\label{temp_zoom_fig}
	\end{figure}

	Fig.\ref{temp_zoom_fig} compares UrbClim (100 m) and downscaled (5 m) air temperature maps, for two selected regions. As depicted in Fig.\ref{temp_zoom_fig}, the downscaling process significantly increased the level of detail of air temperature variations. Additionaly, the impact of urban features and morphology on air temperature variation has been illustrated in achieved high resolution map. This figure consists of three subplots, displaying the topographic map of different areas in Amsterdam along with air temperature maps at 5 m and 100 m resolutions. It shows the higher resolution of air temperature variations and local air temperature patterns (5 m) compared to the 100 m resolution. The figure illustrates the impact of urban features on air temperature in two different regions: the area adjacent to a water channel (A), the city center (areas with high building density) (B). In area A, the downscaled map reveals the river path  with higher clarity, and the effect of buildings on temperatue increase is observable. In area B, the influence of buildings, streets, and water canals is visible with better resolution.

	\section{Discussion}\label{sec_6}
	\subsection{\textcolor{red}{Evaluation of Downscaled Air Temperature Maps}}
	To evaluate the performance of the developed framework, downscaled air temperature maps were created in three modes (average, minimum, and maximum daily) as samples for January 17-19, May 4-6, July 18-20, and October 24-26, 2017. These dates represent the coldest days of winter (January), the hottest days of summer (July), and moderate spring and autumn temperatures (May and October). This evaluation ensures the framework's performance across various air temperature conditions. It should be mentioned that, for better comparison, a raster display of downscaled average, minimum, and maximum daily air temperatures for selected days was provided with the same color bar (Figures S4-S6 in the supplementary). As shown in the maps, air temperature variations in the downscaled air temperature maps are consistent with the season changes.
	
\begin{figure}[!tb]
		\centering
		{\includegraphics[width=\linewidth]{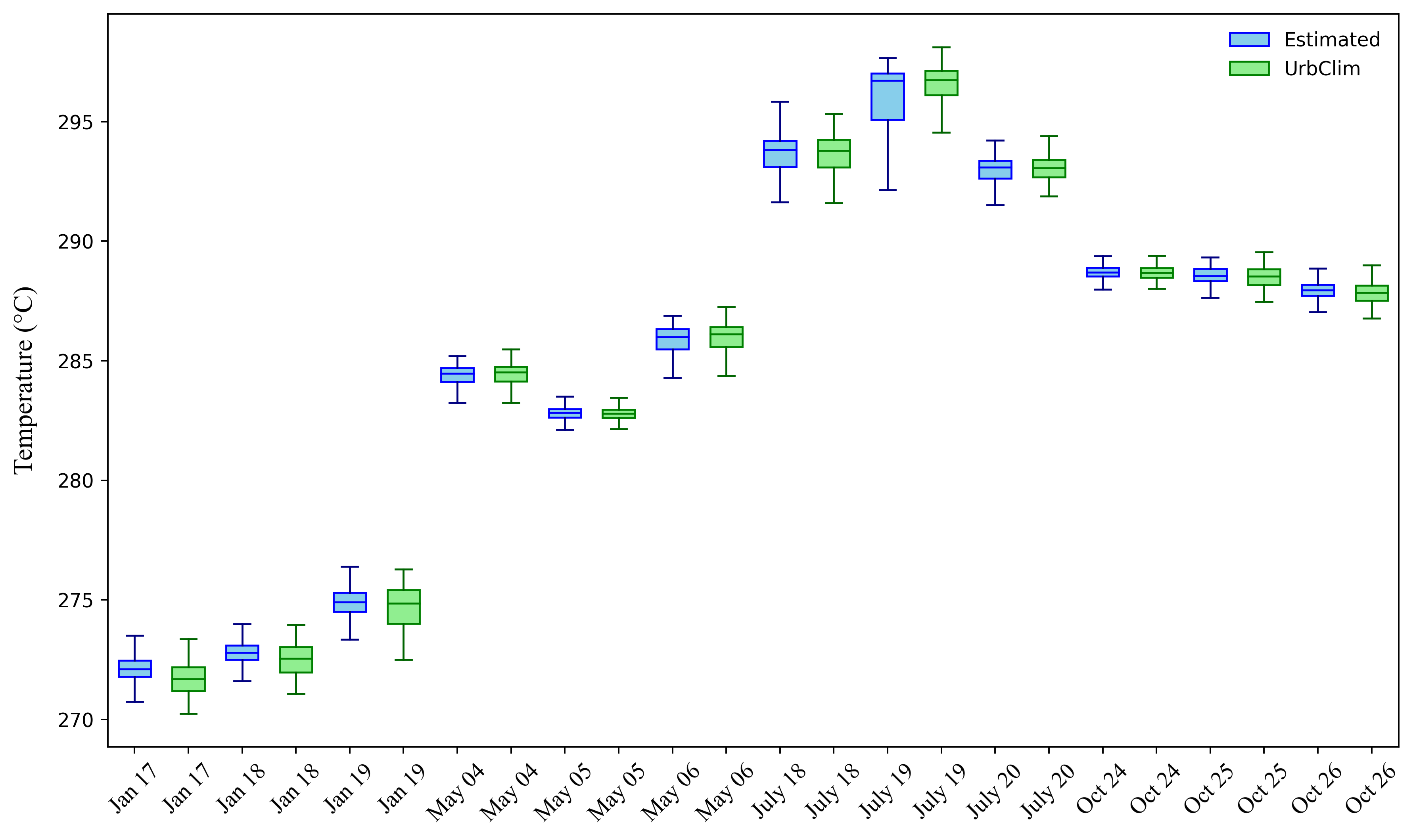}}
		\vspace{0.05cm}
		\caption{Comparison of downscaled air temperature estimations vs. UrbClim model-derived air temperature in the selected days of different seasons for the year of 2017.}
		\label{boxplot_avg_fig}
	\end{figure}

	Due to low variability of air temperature changes across different locations, map alone can’t demonstrate downscaling performance. Therefore, 24 box plots of estimated average temperatures at a 5 m resolution by the LightGBM model for 12 different days were plotted alongside UrbClim model values (100 m resolution) in Fig.\ref{boxplot_avg_fig}. Similarly, box plots for minimum and maximum air temperatures are shown in Figures S7 and S8. The box plots provides a comprehensive comparison between downscaled temperatures and corresponding UrbClim temperatures, and it allows for evaluating the model’s performance in different seasons. By analyzing the box plots, it can be observed that generally the downscaled temperature values are consistent with the UrbClim values, indicating the proposed framework’s ability to downscale air temperatures with reasonable accuracy. However, due to the increased level of detail and more precise representation of air temperature variations, there are differences between these two estimates. The median values of the estimated air temperature are very close to the median values of the UrbClim model. The interquartile range (IQR) values are also very close to each other for most days. However, for some days (January), the IQR value for the UrbClim model is higher, indicating that the UrbClim temperature has higher variability on those days. Also, the downscaling accuracy is higher for moderate seasons (May and October) and better for summer than winter. However, model showed lower performance on July 19 (summer), the large IQR for the downscaled temperature indicates greater spatial variability within the downscaled datasets. While the medians of two models are the same, indicating that on average the downscaled air temperature capture the central tendency of the air temperature distribution represented by the UrbClim model, the lower minimum air temperature in the downscaled version suggests that the downscaled model is predicting cooler air temperatures for certain locations compared to the UrbClim model.

\begin{figure}[!tb]
		\centering
		\begin{center}
			{\includegraphics[width=0.7\linewidth]{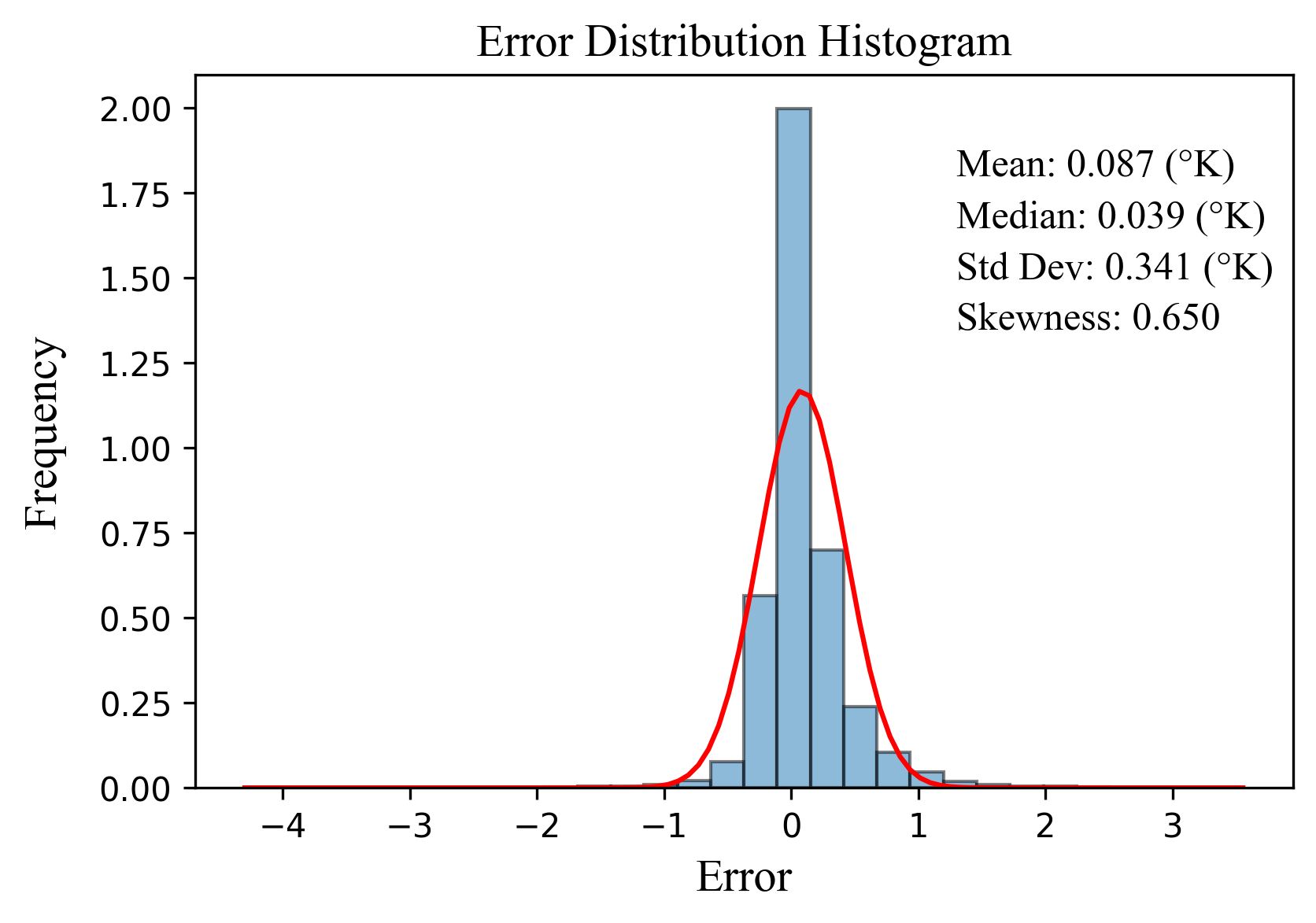}}
			\vspace{0.05cm}
			\caption{Histogram of air temperature estimation error for test points}
			\label{ErrorDist_fig}
		\end{center}
	\end{figure}

	To investigate the range of air temperature variations in more detail, Fig.\ref{ErrorDist_fig} depicts a distribution plot of the estimated air temperature errors for the test points. According to Fig.\ref{ErrorDist_fig}, the error distribution is approximately symmetric and similar to the normal distribution, with low variability (standard deviation 0.341°K) and slight right skewness (skewness 0.650), indicating that the downscaling is capable of estimating air temperatures with high accuracy. The mean error of the data is equal to 0.087°k, and the median is 0.039°k, indicating low bias in the estimated values compared to the UrbClim model. The positive mean and median suggest a slight overestimation bias in the errors. The kurtosis is positive, meaning that its peak is higher than the normal distribution, indicating most errors have values close to zero. Overall, it can be concluded that the range of downscaled temperature variations is almost consistent with the range of air temperature variations in the UrbClim model. In other words, the proposed framework, while increasing the resolution of air temperature estimation, preserves the overall trend and range of air temperature variations.

	According to Fig.\ref{TempMap_fig}.a, the downscaled air temperature map clearly illustrates the relationship between air temperature and the employed predictive features. Moreover, the results obtained from the correlation analysis between input parameters and the predicted air temperature at a 5 m resolution by the LightGBM model confirm a strong correlation between input feature and the predicted air temperature in different urban areas. It is worth noting that initial assessments showed a linear correlation (on average 0.85) between input parameters and the predicted air temperature based on a linear regression model. However, high error in linear model (RMSE 2.534°K, MAE 1.994°K), highlights the need for advanced models such as LightGBM for more accurate air temperature estimation. Analysis also demonstrates a high correlation between RH and the predicted air temperature (-0.91), indicating that increased in air humidity decreases air temperature. According to Fig.\ref{features_fig}, generally, areas closer to water bodies and larger rivers have higher RH, while RH decreases in the city center as the distance from large rivers increases. Additionally, in the city center, the area of walls (WALL), building heights (H), and built-up area (BR) increase compared to the outskirts of the city, and due to the positive and almost equal correlation of all three parameters (0.39, 0.38, and 0.39, respectively), the air temperature increases (see Fig.\ref{TempMap_fig}.a). Furthermore, a high correlation (0.82) exists between wind speed and RH. In cases where the wind passes over water or other areas with high water vapor concentrations (such as channels), higher wind speeds can increase RH and subsequently reduce air temperature.
	
	As shown in Fig.\ref{TempMap_fig}, the details of air temperature variation patterns have increased after the downscaling process. This increase in resolution is attributed to the use of high-resolution features (5 m) as input for the downscaling process. Moreover, the details recorded in the results are consistent with real-world phenomena. For instance, as depicted in Fig.\ref{temp_zoom_fig}, two sample areas have been selected to investigate further the effect of downscaling on increasing details and resolution of air temperature maps. According to Fig.\ref{temp_zoom_fig}.A, in the 5 m resolution map, the air temperature variation patterns in water channels have become more detailed. The more precise measurement of features such as distance to water channels at 5 m resolution has effectively captured the air temperature variation details in the vicinity of water bodies and major rivers compared to the UrbClim model. Additionally, in the city center, the footprint of linear objects such as water channels in air temperature reduction and major streets as impervious surfaces in an increase of air temperature are observable in the 5 m map. In Fig.\ref{temp_zoom_fig}.B, the effect of green spaces within the city center (between buildings) on air temperature variation in the initial map (100 m) is coarser, whereas, thanks to downscaling, the effects of green spaces in the downscaled map have become more accurate. Overall, by comparing the initial air temperature map obtained from the UrbClim model (100 m) with the downscaled temperature map with better resolution (5 m), it is observed that the details of air temperature variations (while preserving the overall pattern) have increased after downscaling, thanks to the increase in the resolution of the input parameters.
	
	In this study, urban morphological features have been extracted from LOD1 building models. However, higher LODs like LOD2 or LOD3, which provide more detailed building information, can accurately capture complex urban features, leading to improved modeling results. Therefore, part of the modeling error may be attributed to simplifying this issue. It is worth noting that while higher LODs offer better details and more accurate representations, they also increase computational complexity and data processing requirements. Hence, a trade-off between detail and computational resources must be considered.
	
	\subsection{Toward a Digital Twin Model Generation of Urban Climate}
\begin{figure}[!htb]
		\centering
		\subfloat{{\includegraphics[width=0.68\linewidth]{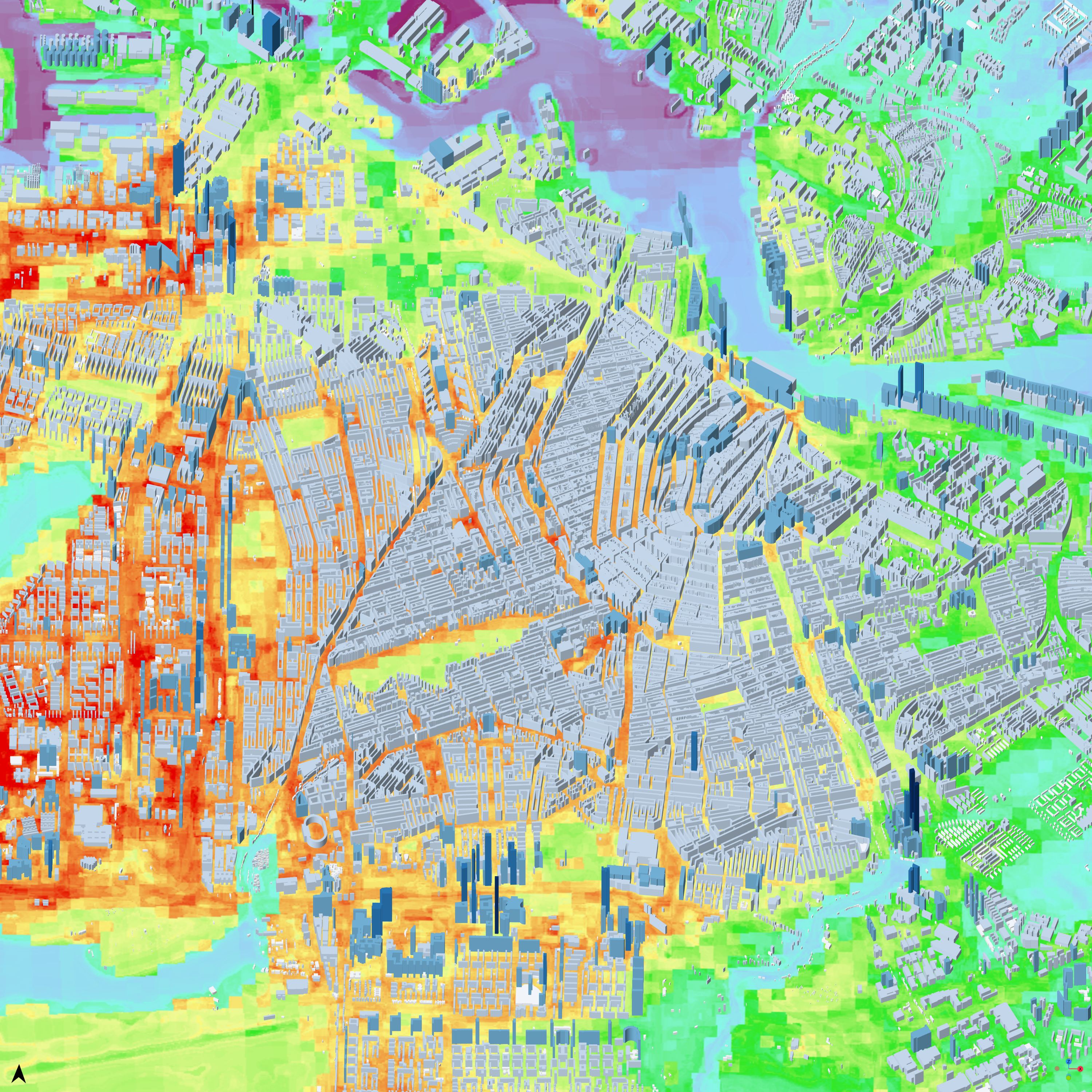}}}%
		\qquad
		\subfloat{{\includegraphics[width=0.22\linewidth]{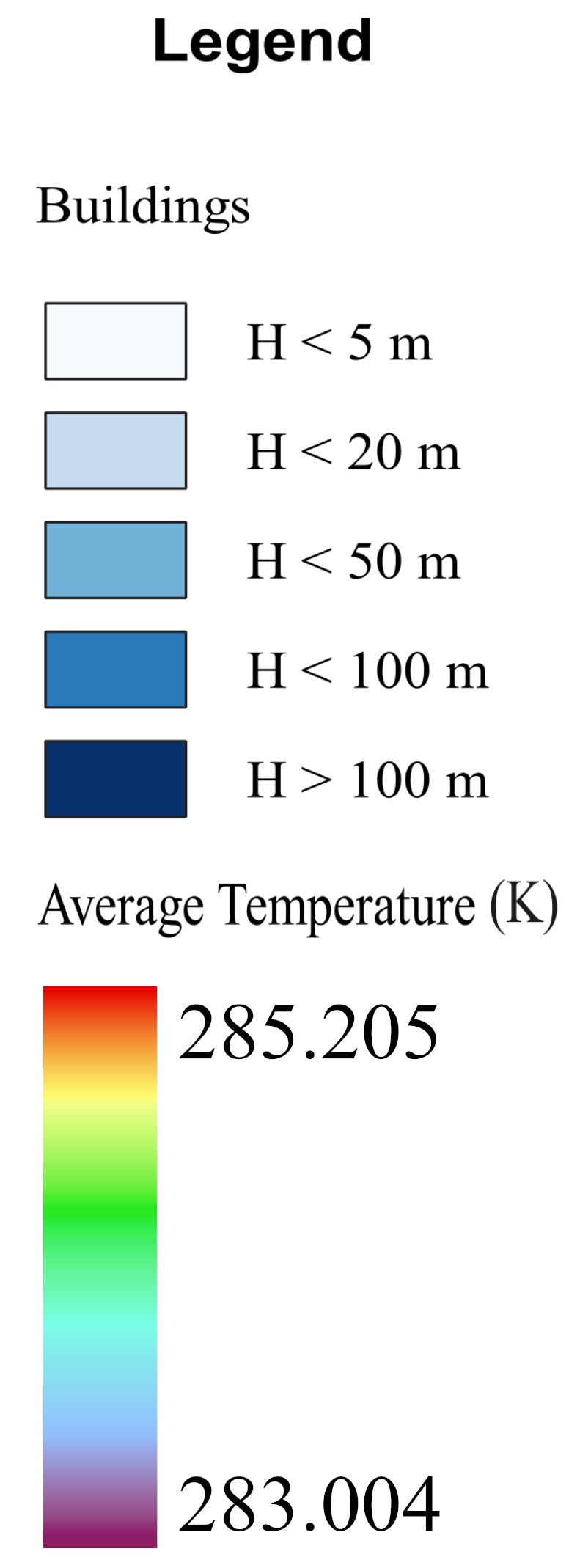} }}%
		\caption{3D building model of Amsterdam city center overlayed with estimated average air temperature on May 4th, 2017}
		\label{3dTemp_fig}
	\end{figure}

	In this section, the potential of using the developed model for generating a digital twin model of Amsterdam's urban climate is discussed. A digital twin is a generally three-dimensional digital representation of a physical object or system in the real world, bridging the gap between the real and digital worlds \citep{RN233}. The city digital twin can help provide a simulated environment and better visualization of the city, and it can be used for design, planning, decision-making, and weather prediction \citep{RN21}. The successful downscaling of air temperature in Amsterdam to a resolution of 5 m using the proposed framework provides opportunities for generating a digital twin of the urban microclimate. By integrating the downscaled temperature data into the 3D model generated from the LiDAR data, a basis is established for creating a digital twin model of the urban microclimate. Fig.\ref{3dTemp_fig} provides an example of this effort to create a digital twin model of the microclimate (air temperature) at the level of Amsterdam. In this figure, it can be observed that areas with taller buildings tend to exhibit higher air temperature values, suggesting a potential influence among other factors. Additionally, lower air temperatures are estimated in open spaces between buildings. In the city center, although there is a higher density of buildings, the traditional fabric and lower height of the buildings, along with the lower DW values, result in relatively lower air temperatures compared to areas with taller buildings.
	
	Using this model, 3D air temperature maps of the city can be generated, and they can be used to identify areas that are more exposed to heat waves and other climate-related hazards. Additionally, other products can be added to the created 3D model. Urban planners and policymakers can utilize this information to implement targeted interventions to mitigate the impact of heat waves and improve the thermal comfort of the city.
	
	\section{Conclusion}\label{sec_7}
	This study presents a framework for downscaling air temperature at the urban scale in Amsterdam using meteorological data from urban climate models such as UrbClim, as well as urban and building morphological features. To achieve this, some of the urban and building morphological features were generated using a 3D model of the buildings. In the first step, building footprints were extracted from LiDAR data using deep learning models. Evaluations showed that the U-Net3+ and Attention U-Net models achieved the highest accuracy in extracting building footprints. Additionally, transfer learning techniques were employed to improve the accuracy of the deep learning models due to limited training data in the study area. After performing post-processing, a 3D model of the buildings was created using the detected footprints and incorporating height information from LiDAR point clouds. Then, urban morphological features such as building heights (H), building coverage (BR), total exterior wall area (WALL), and canopy coverage (CCR) were extracted from the achived 3D model. In addition to these features, other features such as sky view factor (SVF), solar radiation (SR), distance to water and parks (DW, DP), and pervious surfaces ratio were extracted from LiDAR and OSM data. These features, along with meteorological data from the UrbClim model, were used for downscaling air temperature at the city level. Various machine learning algorithms were evaluated for this purpose, and the results showed that the LightGBM model had the highest accuracy. In the end, the developed framework was used to generate a downscaled temperature map at a resolution of 5 m for the city of Amsterdam. The evaluation of the generated maps demonstrated that the proposed method had high capability in detecting and visualizing local air temperature patterns at the microclimate level. These results are valuable for better understanding the urban microclimate and its impacts on urban planning, energy consumption, and human comfort. It is worth mentioning that although the proposed framework utilizes UrbClim data as the base data for downscaling air temperature, it has the potential to be used and generalized to the outputs of other models, such as ENVI-met, enabling downscaling of the outputs of these physically-based models at lower resolutions to reduce computational costs. Furthermore, considering the generated 3D city model, it can be a basis for generating a digital twin of urban climate. In future research, more advanced deep learning algorithms could be explored for air temperature downscaling purposes.
	
	\section*{Acknowledgement}
	We would like to present our acknowledgments to Mr. Jonas Nelson, from Municipality of Taby, Sweden, for generously sharing his FME workspace for modelling trees, \href{https://www.openstreetmap.org/}{OpenStreetMap} for providing the essential geospatial data, \href{https://ahn.nl/}{AHN} for providing the Amsterdam LiDAR data,  \href{https://www.knmi.nl/over-het-knmi/about}{KNMI} for providing ground meteorological measurements,  \href{https://cds.climate.copernicus.eu/}{Copernicus Climate Data Store} for providing UrbClim model data used in this research. We also would like to express our appreciation for the review efforts of the two anonymous reviewers.

	\bibliographystyle{model1-num-names}
	\bibliography{ReferenceLibrary.bib}
	
	\begin{appendices}

	\counterwithin{figure}{section}
	\counterwithin{table}{section}
		\section{Detailed Structures of Deep Semantic Segmentation Models}\label{appendix_dl_fig}
\begin{figure}[!h]
				\centering
				{\includegraphics[width=0.95\linewidth]{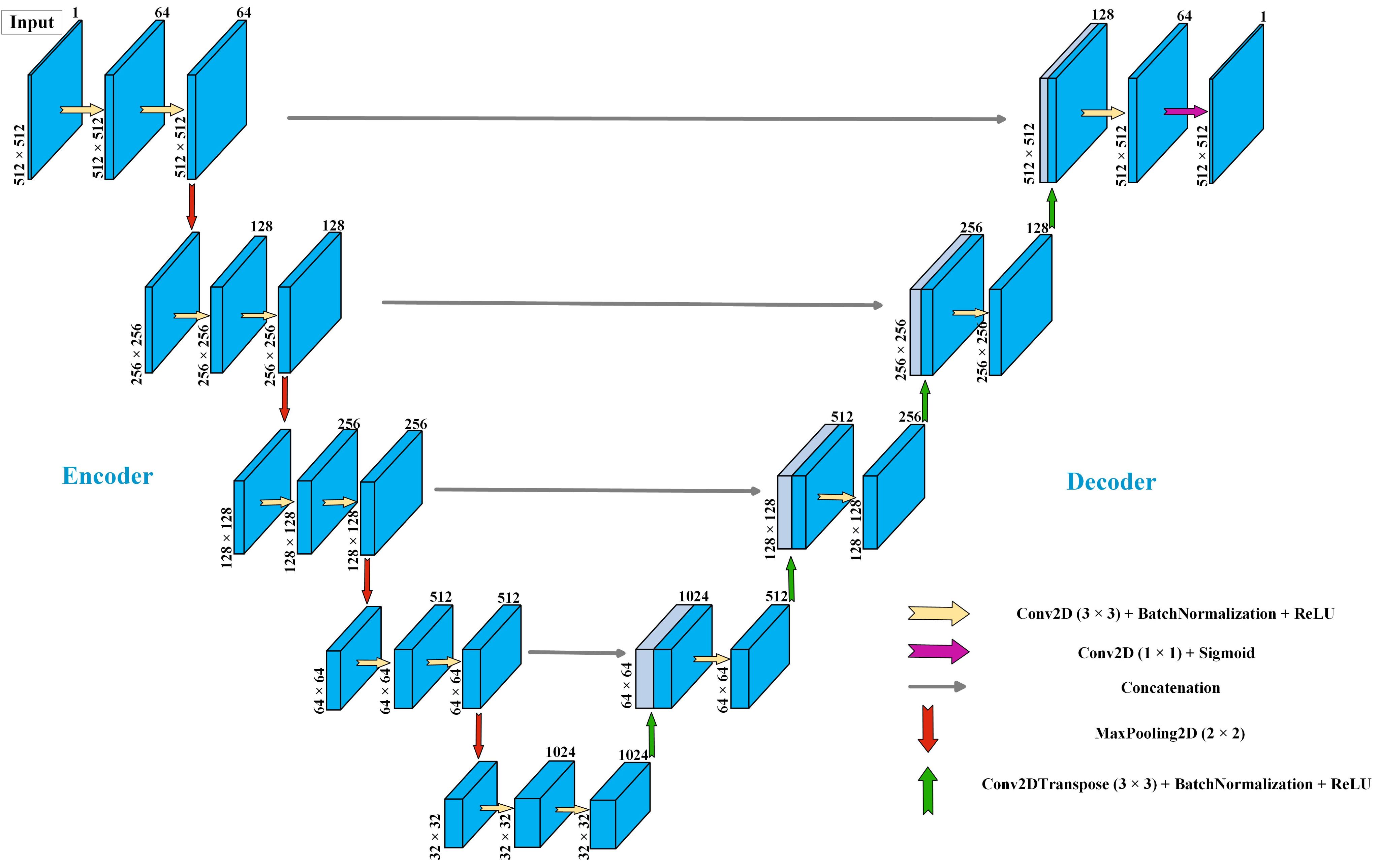}}
				\vspace{0.05cm}
				\caption{Architecture of U-Net model used in this study for building footprint extraction}
				\label{unet_fig}
			\end{figure}

\begin{figure}[!h]
				\centering
				{\includegraphics[width=0.95\linewidth]{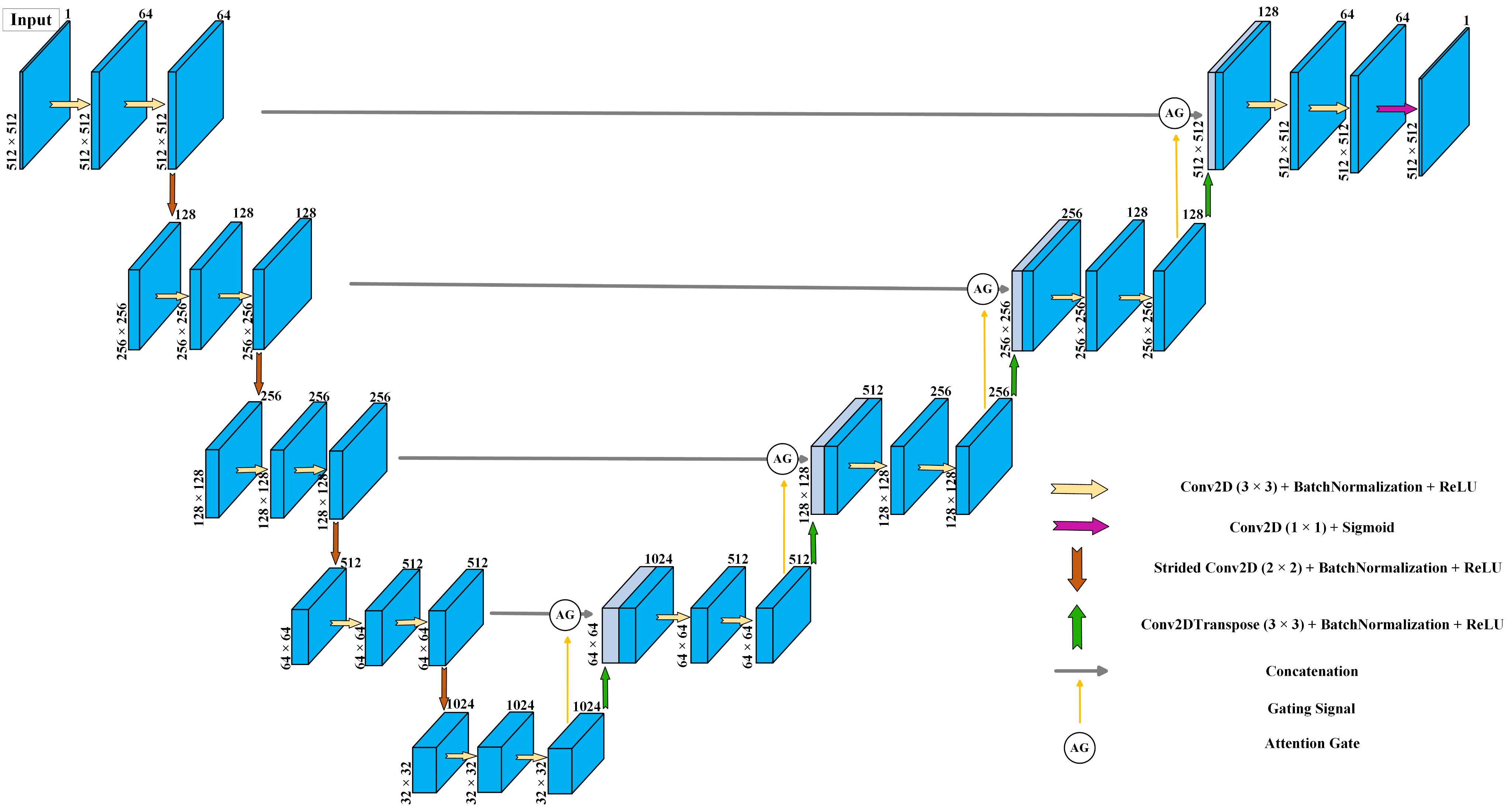}}
				\vspace{0.05cm}
				\caption{Architecture of Attention U-Net used in this study for building footprint extraction}
				\label{attention_fig}
			\end{figure}

\begin{figure}[!h]
				\centering
				{\includegraphics[width=\linewidth]{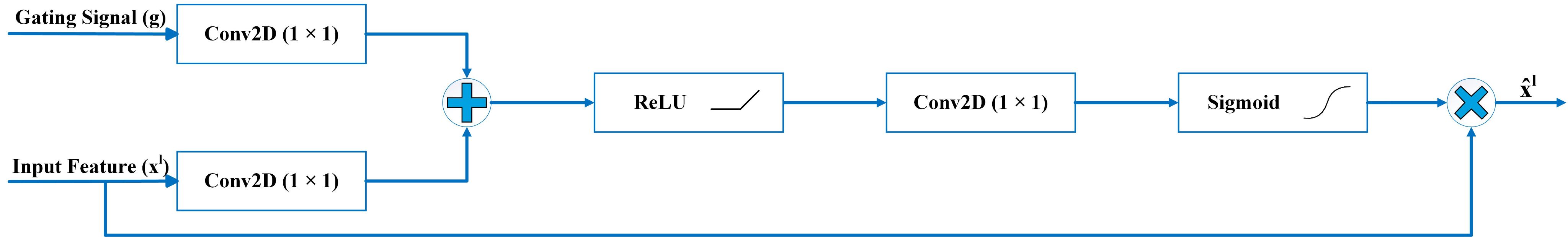}}
				\vspace{0.05cm}
				\caption{Schematic visualization of Attention Gate (AG) mechanism, it is performed right before the concatenation operation to enable the model to focus on the most important features by scaling the input features with computed attention coefficients ($\alpha$) in AG.}
				\label{AG_fig}
			\end{figure}

\begin{figure}[!h]
				\centering
				{\includegraphics[width=\linewidth]{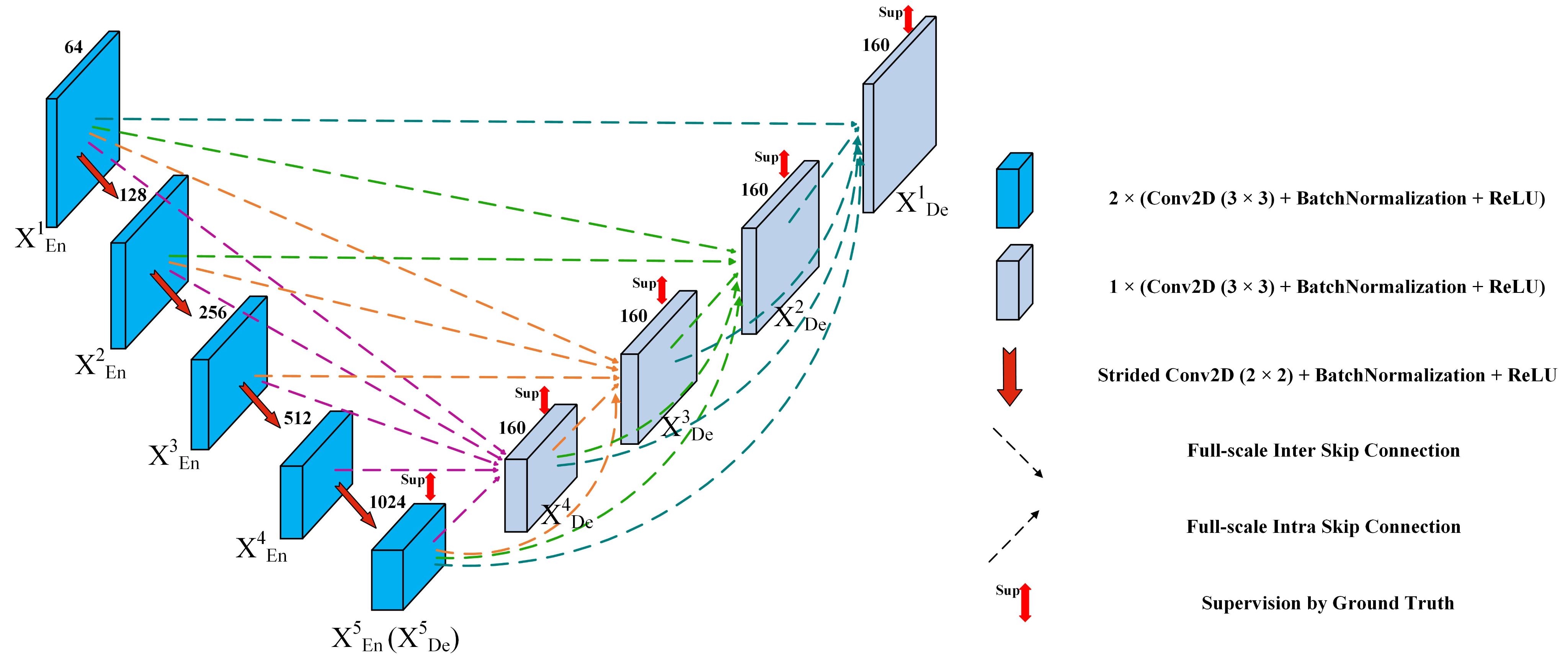}}
				\vspace{0.05cm}
				\caption{Architecture of U-Net3+ used in this study for building footprint extraction}
				\label{unet3p_fig}
			\end{figure}

\begin{figure}[!h]
				\centering
				{\includegraphics[width=\linewidth]{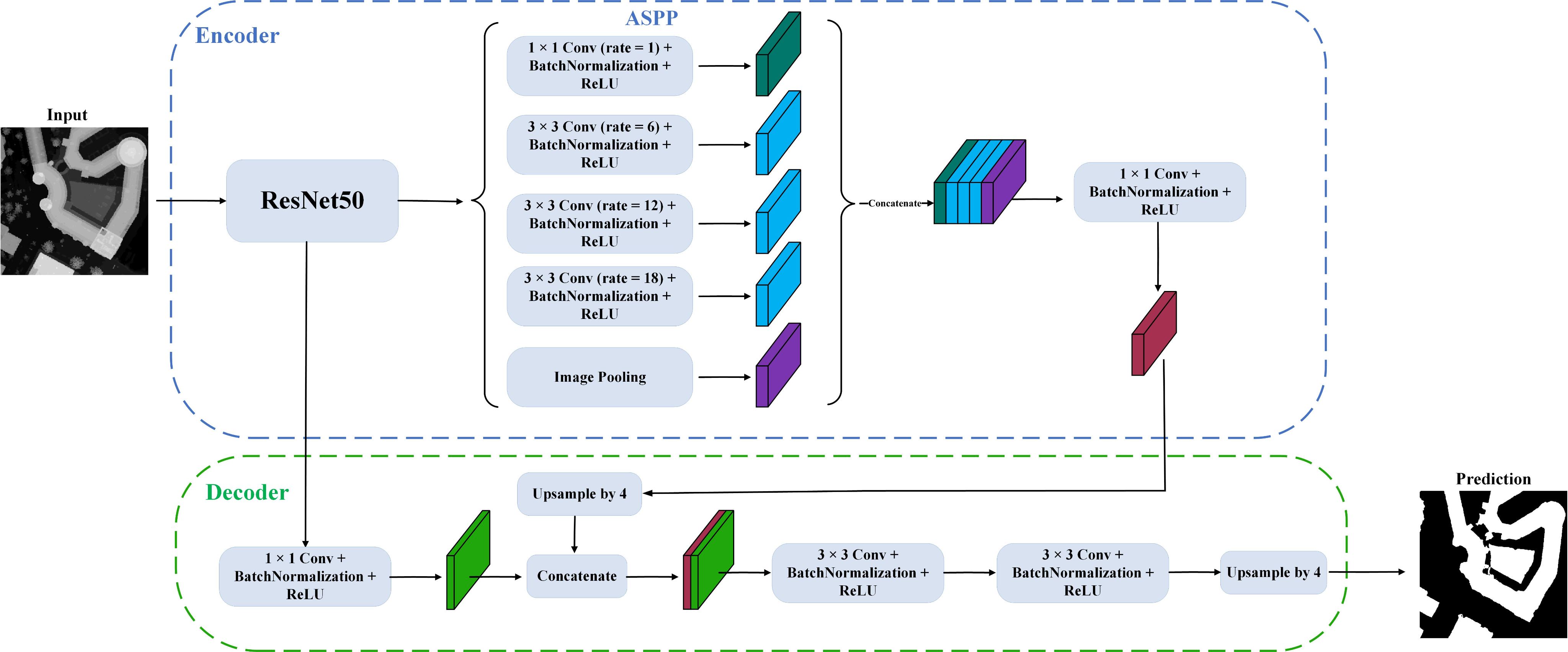}}
				\vspace{0.05cm}
				\caption{Architecture of DeepLabV3+ used in this study for building footprint extraction}
				\label{deeplab_fig}
			\end{figure}

			Figures. \ref{unet_fig}-\ref{deeplab_fig} exhibit the designed structures of deep semantic models used in this study for building foot print extraction. 
		
		\section{Morphological Feature Extraction} \label{appendix_features}
		This section provides more details on calculation of morphological features introduced in Section \ref{sec_4.4}.
		
		\begin{itemize}
			\item{Sky View Factor (SVF):}
		\end{itemize}
		The SVF was computed using QGIS \textcolor{red}{and the UMEP plugin}. The input to the software was the DSM of the study area, and the SVF parameter was generated according to the resolution of the input DSM. Then, the output raster was resampled to the target resolution of 5 m. Finally, the values obtained in the influence area were averaged.

		\begin{itemize}
			\item{Average Building Height (H):}
		\end{itemize}
		\textcolor{red}{To calculate the building heights parameter, the heights of all buildings within the influence area were averaged. Each building's height was determined from the median height of the LiDAR points within its footprint during the creation of the 3D model.}
		
		\begin{itemize}
			\item{Building Coverage (BR):}
		\end{itemize}
		
		To calculate BR, the intersection between the building footprints and the influence area was determined. The ratio of the sum of the intersected areas to the total area of the influence area (e.g., 50 m$\times$50 m=11025 $m^2$) is then calculated:
		\begin{equation}
			\text{BR} = \frac {\sum_{i}^{N}a_i}{A_T}, \label{bdg_eq}
		\end{equation}
		where i denotes the building footprint, $a_i$ represents the intersected area of each building, and $A_T$ is the total area of the influence area.
		
		\begin{itemize}
			\item{Total Exterior Wall Area (WALL):}
		\end{itemize}
		
\begin{figure}[!tb]
			\centering
			{\includegraphics[width=0.6\linewidth]{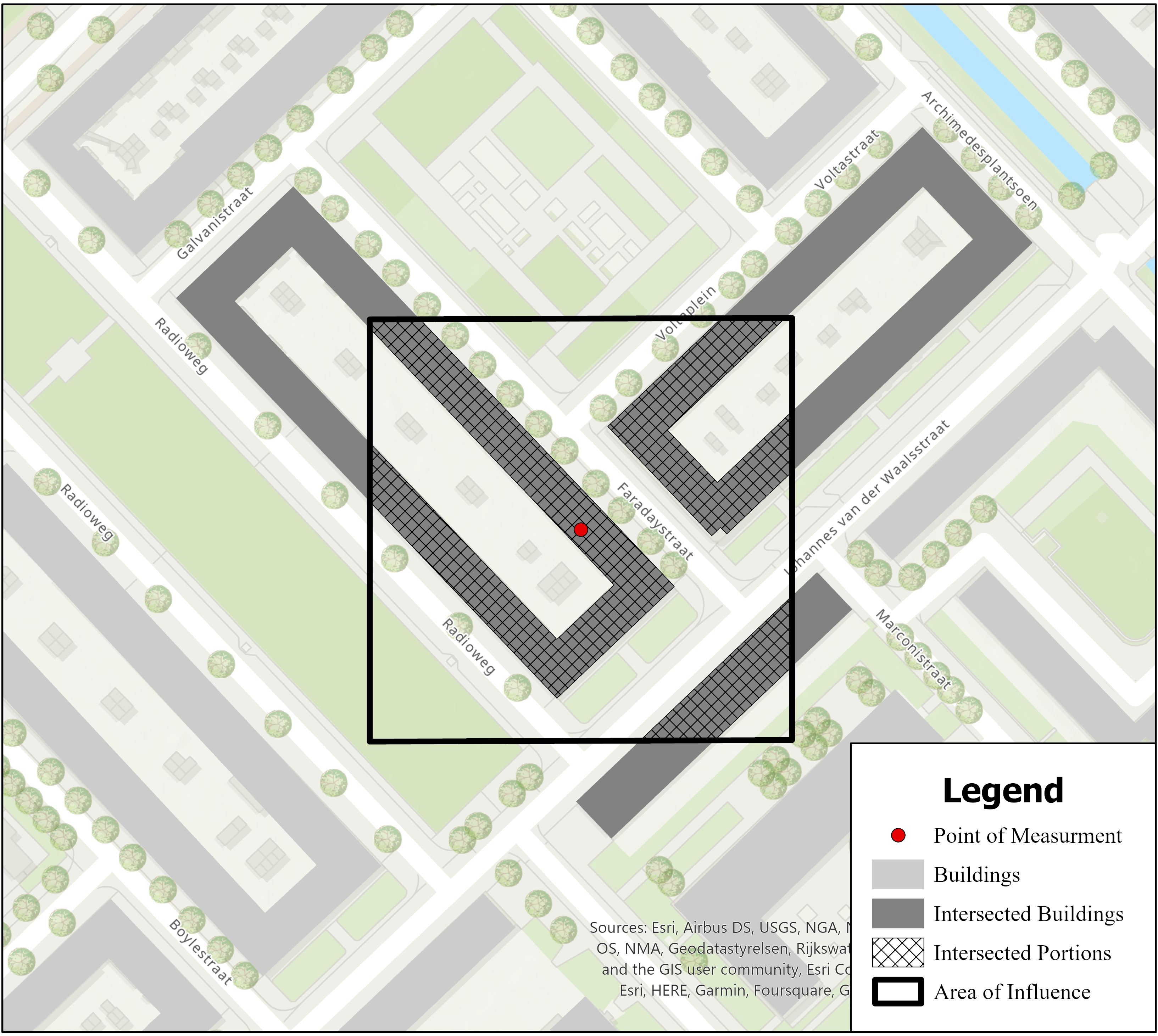}}
			\vspace{0.05cm}
			\caption{Intersection between area of influence and buildings for calculation of total exterior wall area.}
			\label{wall_fig}
		\end{figure}

		In determining the exterior wall area (WALL), walls were separated from other building components in the CityGML model. \textcolor{red}{First, the different components of the buildings in the CityGML model were separated. By calculating the slope of each segment, the walls were identified: a slope of 90° corresponded to walls, and a slope of zero corresponded to the roof.} The total wall area was then calculated for each building. \textcolor{red}{Since a building might not have been fully within the considered influential radius} (e.g., Fig.\ref{wall_fig}), a weight was assigned to the wall area by considering the ratio of the intersected building area within the influential radius ($a_i$) to the total area of the building ($A_i$). The exterior wall area for each building in the influence area was the product of the obtained weight and the total area of the walls of that building. Finally, the total area of the exterior walls of the existing buildings in the influence area was calculated according to eq. \ref{wall_eq}:
		\begin{equation}
			\text{WALL} = \sum_{i}(\frac{a_i}{A_i} \times \sum_{i}Wall_i), \label{wall_eq}
		\end{equation}
		where i denotes the building footprint, and $Wall_i$ represents the area of the exterior walls of the building i in the influential region.
		
		\begin{itemize}
			\item{Distance to Parks (DP):}
		\end{itemize}
		
		For calculating DP, first, the Euclidean distance between each pixel and the nearest park was calculated using the OSM park layer. The resulting raster had a resolution of 5 m, aligning with the target resolution for downscaling air temperature. This raster specified the DP value in each cell, representing the distance of each location to the nearest park. Finally, the average DP values within the influence area were calculated.
		
		\begin{itemize}
			\item{\textcolor{red}{Distance to Waters (DW):}}
		\end{itemize}
		
		The distance to water (DW) was computed using the water layer of OSM, following the same methodology as for DP. The DW raster, with a 5 m resolution, depicted the distance of each pixel to the nearest water body.
		
		\begin{itemize}
			\item{Canopy Cover Ratio (CCR):}
		\end{itemize}
		
\begin{figure}[t]
			\centering
			{\includegraphics[width=0.6\linewidth]{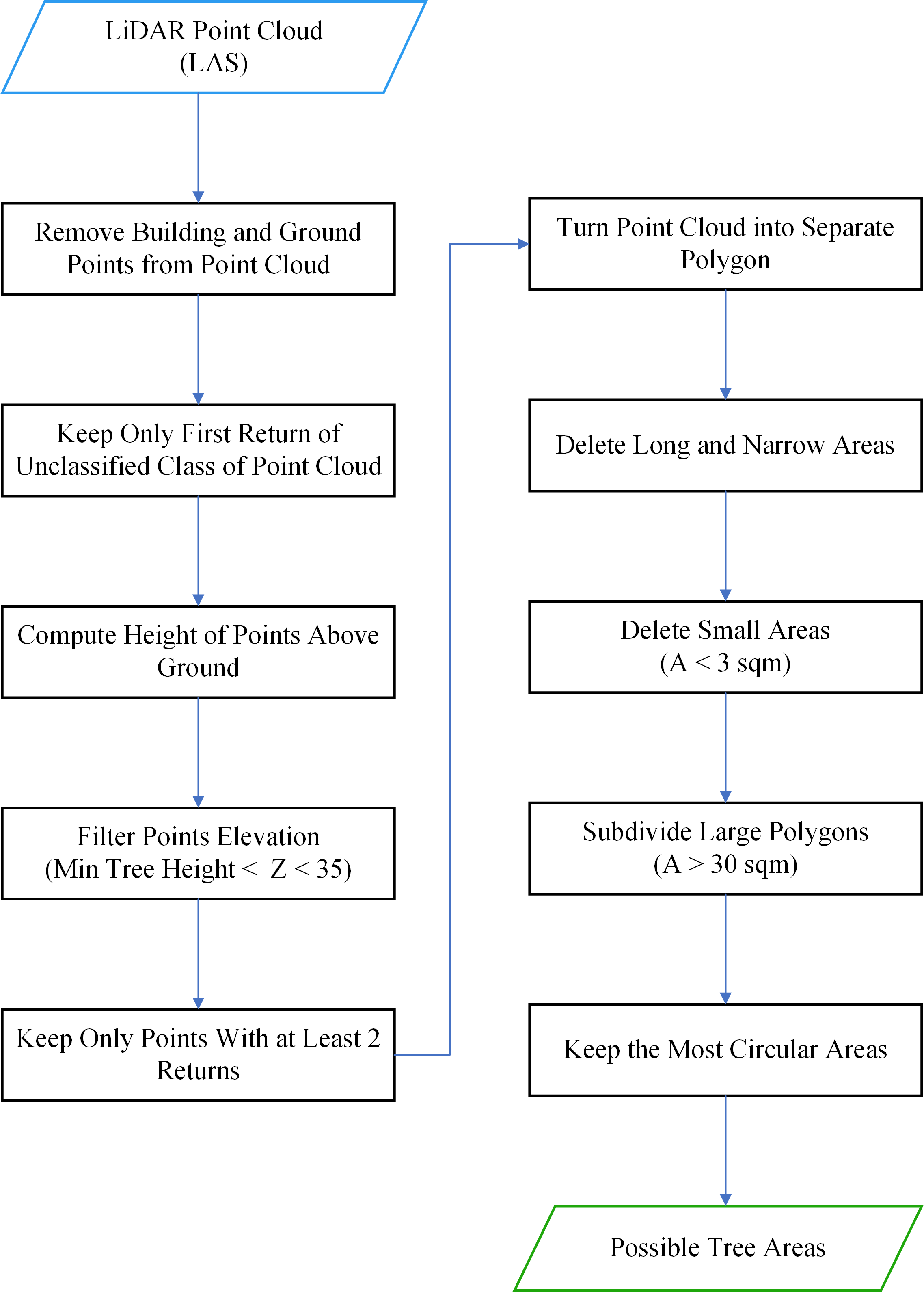}}
			\vspace{0.05cm}
			\caption{Process of extracting tree canopy cover from LiDAR point cloud data}
			\label{tcr_flowchart}
		\end{figure}

		To calculate CCR index, according to the process shown in Fig.\ref{tcr_flowchart}, points belonging to trees were identified from the LiDAR point cloud. These points were identified based on features such as height and LiDAR reflectance. In this way, the tree canopy was determined \citep{RN227}. If the LiDAR point cloud data has already been classified, the high vegetation class identifies tree points. Since the LiDAR data used in this study did not have high vegetation class information, it was necessary to extract trees from the LiDAR point cloud according to the process in Fig.\ref{tcr_flowchart}. According to Fig.\ref{tcr_flowchart}, first, points related to the ground and buildings were removed from the initial LiDAR point cloud. The points belonging to buildings were selected  by the footprints extracted using U-Net3+ in this research, and then removed from the initial LiDAR point cloud. Then, a one-meter buffer was considered around the building points to remove artificial elements such as windows. From the remaining points, only the LiDAR first return points, representing the top of tree canopy height, were retained. Next, the heights were normalized to obtain the heights above ground. Then, points with heights less than a threshold (defaulted to 4 m as the minimum tree height) and points with heights greater than 35 m (noise) were removed. Typically, LiDAR hits trees with multiple returns compared to other surfaces, so points with at least two returns were considered from the remained points. In the next step, the remaining points were converted into polygons. Among the polygons, long and narrow polygons that might have belonged to bridges or power lines were removed. After calculating the area of each polygon, polygons with an area less than three square meters were also removed (which could be related to streetlights), and polygons with an area larger than 30 square meters, likely indicating multiple adjacent trees, were subdivided. In the next stage, the circularity of each polygon was calculated. For polygons with an area larger than 30 square meters, a circularity factor greater than 0.2 was acceptable, and for other polygons, a circularity factor greater than 0.7 was acceptable. Finally, the obtained tree polygons were saved in the desired format (e.g., shapefile). 
		
		After saving the tree polygons in desired format, the vector file representing tree canopy was converted to a raster, where a value of 1 represented tree pixels, and a value of 0 represented other pixels. Finally, the ratio of the area with tree cover to the total area of the influence area was considered as the final value of the CCR variable as below:
		\begin{equation}
			\text{CCR} = \frac{\sum t \times a_{pix}}{A_T}, \label{tcr_eq}
		\end{equation}
		where t is a binary variable, which is equal to 1 if the pixel contains a tree and 0 otherwise. $a_{pix}$ represents the area of a pixel, and $A_T$ denotes the total area of the influence area.
		
		\begin{itemize}
			\item{Pervious Surface Ratio (P):}
		\end{itemize}
		
		For calculating pervious surface ratio (P), the input data were merged and converted to a raster, in which a value of 1 represented pixels related to pervious surfaces, and a value of 0 illustrated other areas. Then, similar to other variables, the average area of pervious surfaces in the influence area was calculated according to relation below:
		\begin{equation}
			\text{P} = \frac{\sum p \times a_{pix}}{A_T}, \label{pervious_eq}
		\end{equation}
		where p is a binary variable, which is equal to 1 if the pixel denotes pervious areas and 0 otherwise. $a_{pix}$ represents the area of a pixel, and $A_T$ denotes the total area of the influence area.
		\begin{itemize}
			\item{\textcolor{red}{Solar Radiation (SR):}}
		\end{itemize}
		\textcolor{red}{The SR variable was calculated as the daily average for four months of January, May, July, and October in 2017 (concurrent with air temperature data) using the Area Solar Radiation tool in ArcGIS Pro 3 software.}
		\begin{itemize}
			\item{\textcolor{red}{Wind Speed (WS):}}
		\end{itemize}
		\textcolor{red}{Generally, wind tends to reduce air temperature as it leads to evaporation, which is a cooling process. It also affects air movement. Wind can mix air and bring cooler air from higher elevations to the surface. However, in some cases, it can increase the air temperature. For example, when the wind blows from a warmer location, it brings warmer air to the new location. 
		Since the goal is to estimate air temperature as an average, minimum, or maximum, wind speed data from the UrbClim model were also calculated in the data preparation stage for the average, minimum, and maximum daily air temperature. Additionally, it's worth noting that the output of UrbClim is at a 100 m resolution, which was resampled using a simple resampling technique (nearest neighbor) to achieve a 5 m resolution.}

		\section{Regression Model Hyperparameter Tuning} \label{appendix_hp_table}
		
		\begin{table}[h]
			\centering \footnotesize
			\caption{Hyperparameter settings of the regression models}
			\begin{tabularx}{\textwidth}{ c |X}
				
				\textbf{Model} & \textbf{Hyperparameters} \\
				\hline
				LightGBM & \begin{itemize}[nosep,leftmargin=*]
					\item Learning Rate = 0.045
					\item Max Depth = 10
					\item No. of Estimators = 3000
					\item No. of Leaves = 28
					\item Boosting Type = gbdt
					\item Colsample Bytree = 0.92
					\item Subsample = 0.22
					\item Reg Lambda = 0.73
				\end{itemize} \\
				\hline
				XGBoost & \begin{itemize}[nosep,leftmargin=*]
					\item Eta = 0.1
					\item Max Depth = 10
					\item No. of Estimators = 1000
					\item Colsample Bytree = 0.75
					\item Subsample = 0.75
					\item Min Child Weight = 5
				\end{itemize} \\
				\hline
				SVR & \begin{itemize}[nosep,leftmargin=*]
					\item Epsilon = 0.1
					\item C = 100
					\item Kernel Type = RBF
				\end{itemize} \\
				\hline
				RF & \begin{itemize}[nosep,leftmargin=*]
					\item Max Depth = 10
					\item No. of Estimators = 500
					\item Max Features = 0.7
				\end{itemize} \\
				\hline
				ET & \begin{itemize}[nosep,leftmargin=*]
					\item Max Depth = 10
					\item No. of Estimators = 1000
					\item Max Features = 0.8
				\end{itemize} \\
				\hline
			\end{tabularx} 
			\label{table_hp}
		\end{table}
		
	\end{appendices}
	
\end{document}